\documentclass[12pt,reqno]{amsart}
\usepackage{graphicx}
\usepackage{amscd}
\usepackage{amsmath}
\usepackage{epsfig}
\usepackage{amsfonts}
\usepackage{amssymb}

\providecommand{\U}[1]{\protect\rule{.1in}{.1in}}
\providecommand{\U}[1]{\protect\rule{.1in}{.1in}}
\allowdisplaybreaks

\theoremstyle{plain}

\newtheorem{lemma}{Lemma}

\newtheorem{remark}{Remark}

\numberwithin{equation}{section}

\input{tcilatex}

\begin{document}
\title[Time Inversion for Oscillators]{The Time Inversion for Modified
Oscillators}
\author{Ricardo Cordero-Soto}
\address{School of Mathematical and Statistical Sciences, Mathematical and
Computational Modeling Center, Arizona State University, Tempe, AZ
85287--1804, U.S.A.}
\email{ricardojavier81@gmail.com}
\author{Sergei K. Suslov}
\address{School of Mathematical and Statistical Sciences, Mathematical and
Computational Modeling Center, Arizona State University, Tempe, AZ
85287--1804, U.S.A.}
\email{sks@asu.edu}
\urladdr{http://hahn.la.asu.edu/\symbol{126}suslov/index.html}
\date{\today }

\begin{abstract}
We discuss a new completely integrable case of the time-dependent Schr\"{o}%
dinger equation in $\boldsymbol{R}^{n}$ with variable coefficients for a
modified oscillator, which is dual with respect to the time inversion to a
model of the quantum oscillator recently considered by Meiler, Cordero-Soto,
and Suslov. A second pair of dual Hamiltonians is found in the momentum
representation. Our examples show that in mathematical physics and quantum
mechanics a change in the direction of time may require a total change of
the system dynamics in order to return the system back to its original
quantum state. Particular solutions of the corresponding nonlinear Schr\"{o}%
dinger equations are obtained. A Hamiltonian structure of the classical
integrable problem and its quantization are also discussed.
\end{abstract}

\maketitle

\section{Introduction}

The Cauchy initial value problem for the Schr\"{o}dinger equation%
\begin{equation}
i\frac{\partial \psi }{\partial t}=H\left( t\right) \psi  \label{in1}
\end{equation}%
for a certain modified oscillator is explicitly solved in Ref.~\cite%
{Me:Co:Su} for the case of $n$ dimensions in $\boldsymbol{R}^{n}.$ When $n=1$
the Hamiltonian considered by Meiler, Cordero-Soto, and Suslov has the form%
\begin{equation}
H\left( t\right) =\frac{1}{2}\left( aa^{\dagger }+a^{\dagger }a\right) +%
\frac{1}{2}e^{2it}a^{2}+\frac{1}{2}e^{-2it}\left( a^{\dagger }\right) ^{2},
\label{in2}
\end{equation}%
where the creation and annihilation operators are defined as in \cite{Flu}:%
\begin{equation}
a^{\dagger }=\frac{1}{i\sqrt{2}}\left( \frac{\partial }{\partial x}-x\right)
,\qquad a=\frac{1}{i\sqrt{2}}\left( \frac{\partial }{\partial x}+x\right) .
\label{in3}
\end{equation}%
The corresponding time evolution operator is found in \cite{Me:Co:Su} as an
integral operator%
\begin{equation}
\psi \left( x,t\right) =U\left( t\right) \psi \left( x,0\right)
=\int_{-\infty }^{\infty }G\left( x,y,t\right) \ \psi \left( y,0\right) \ dy
\label{in4}
\end{equation}%
with the kernel (Green's function or propagator) given in terms of
trigonometric and hyperbolic functions as follows%
\begin{eqnarray}
G\left( x,y,t\right) &=&\frac{1}{\sqrt{2\pi i\left( \cos t\sinh t+\sin
t\cosh t\right) }}  \label{in5} \\
&&\times \exp \left( \frac{\left( x^{2}-y^{2}\right) \sin t\sinh
t+2xy-\left( x^{2}+y^{2}\right) \cos t\cosh t}{2i\left( \cos t\sinh t+\sin
t\cosh t\right) }\right) .  \notag
\end{eqnarray}

It is worth noting that the time evolution operator is known explicitly only
in a few special cases. An important example of this source is the forced
harmonic oscillator originally considered by Richard Feynman in his path
integrals approach to the nonrelativistic quantum mechanics \cite{FeynmanPhD}%
, \cite{Feynman}, \cite{Feynman49a}, \cite{Feynman49b}, and \cite{Fey:Hib};
see also \cite{Lop:Sus}. Since then this problem and its special and
limiting cases were discussed by many authors; see Refs.~\cite{Beauregard},
\cite{Gottf:T-MY}, \cite{Holstein}, \cite{Maslov:Fedoriuk}, \cite{Merz},
\cite{Thomber:Taylor} for the simple harmonic oscillator and Refs.~\cite%
{Arrighini:Durante}, \cite{Brown:Zhang}, \cite{Holstein97}, \cite{Nardone},
\cite{Robinett} for the particle in a constant external field and references
therein. Furthermore, in Ref.~\cite{Cor-Sot:Lop:Sua:Sus} the time evolution
operator for the one-dimensional Schr\"{o}dinger equation (\ref{in1}) has
been constructed in a general case when the Hamiltonian is an arbitrary
quadratic form of the operator of coordinate and the operator of linear
momentum. In this approach, the above mentioned exactly solvable models,
including the modified oscillator of \cite{Me:Co:Su}, are classified in
terms of elementary solutions of a certain characteristic equation related
to the Riccati differential equation.

In the present paper we find the time evolution operator for a
\textquotedblleft dual\textquotedblright\ time-dependent Schr\"{o}dinger
equation of the form%
\begin{equation}
i\frac{\partial \psi }{\partial \tau }=H\left( \tau \right) \psi ,\qquad
\tau =\frac{1}{2}\sinh \left( 2t\right)  \label{in6}
\end{equation}%
with another \textquotedblleft exotic\textquotedblright\ Hamiltonian of a
modified oscillator given by%
\begin{equation}
H\left( \tau \right) =\frac{1}{2}\left( aa^{\dagger }+a^{\dagger }a\right) +%
\frac{1}{2}e^{-i\arctan \left( 2\tau \right) }a^{2}+\frac{1}{2}e^{i\arctan
\left( 2\tau \right) }\left( a^{\dagger }\right) ^{2}.  \label{in7}
\end{equation}%
We show that the corresponding propagator can be obtain from expression (\ref%
{in5}) by interchanging the coordinates $x\leftrightarrow y.$ This implies
that these two models are related to each other with respect to the
inversion of time, which is the main result of this article.

The paper is organized as follows. In section~2 we derive the propagators
for the Hamiltonians (\ref{in2}) and (\ref{in7}) following the method of
\cite{Cor-Sot:Lop:Sua:Sus} --- expression (\ref{in5}) was obtain in \cite%
{Me:Co:Su} by a totally different approach using $SU\left( 1,1\right) $%
-symmetry of the $n$-dimensional oscillator wave functions and the
Meixner--Pollaczek polynomials. Another pair of completely integrable dual
Hamiltonians is also discussed here. The \textquotedblleft
hidden\textquotedblright\ symmetry of quadratic propagators is revealed in
section~3. The next section is concerned with the complex form of the
propagators, which unifies Green's functions for several classical models by
geometric means. In section~5 we consider the inverses of the corresponding
time evolution operators and its relation with the inversion of time. A
transition to the momentum representation in section~6 gives the reader a
new insight on the symmetries of the quadratic Hamiltonians under
consideration together with a set of identities for the corresponding time
evolution operators. The $n$-dimensional case is discussed in sections~7 and
8. Particular solutions of the corresponding nonlinear Schr\"{o}dinger
equations are constructed in section~9. The last section is concerned with
the ill-posedness of the Schr\"{o}dinger equations. Three Appendixes at the
end of the paper deal with required solutions of a certain type of
characteristic equations, a quantum Hamiltonian transformation, and a
Hamiltonian structure of the characteristic equations under consideration,
respectively.

As in \cite{Cor-Sot:Lop:Sua:Sus}, \cite{Me:Co:Su} and \cite{Sua:Sus}, we are
dealing here with solutions of the time-dependent Schr\"{o}dinger equation
with variable coefficients. The case of a corresponding diffusion-type
equation is investigated in \cite{Sua:Sus:Vega}. These exactly solvable
models are of interest in a general treatment of the nonlinear evolution
equations; see \cite{Cann}, \cite{CarlesBook}, \cite{Caz}, \cite{Caz:Har},
\cite{Fadd:Takh}, \cite{Lad:Sol:Ural}, \cite{Meln:Filin}, \cite{Tao} and
\cite{Ban:Vega}, \cite{Car}, \cite{Carl}, \cite{Carle}, \cite{Carles}, \cite%
{Carles:Miller}, \cite{Carles:Nakamura}, \cite{Ken:Pon:Veg}, \cite%
{Naibo:Stef}, \cite{Per-G:Tor:Mont}, \cite{Rod:Schlag}, \cite{Roz}, \cite%
{Schlag}, \cite{Staf:Tat} and references therein. They facilitate, for
instance, a detailed study of problems related to global existence and
uniqueness of solutions for the nonlinear Schr\"{o}dinger equations with
general quadratic Hamiltonians. Moreover, these explicit solutions can also
be useful when testing numerical methods of solving the Schr\"{o}dinger and
diffusion-type equations with variable coefficients.

\section{Derivation of The Propagators}

The fundamental solution of the time-dependent Schr\"{o}dinger equation with
the quadratic Hamiltonian of the form%
\begin{equation}
i\frac{\partial \psi }{\partial t}=-a\left( t\right) \frac{\partial ^{2}\psi
}{\partial x^{2}}+b\left( t\right) x^{2}\psi -i\left( c\left( t\right) x%
\frac{\partial \psi }{\partial x}+d\left( t\right) \psi \right)  \label{sol1}
\end{equation}%
in two interesting special cases, namely,%
\begin{equation}
a=\cos ^{2}t,\quad b=\sin ^{2}t,\quad c=2d=\sin \left( 2t\right)
\label{sol2}
\end{equation}%
and%
\begin{equation}
a=\cosh ^{2}t,\quad b=\sinh ^{2}t,\quad c=2d=-\sinh \left( 2t\right) ,
\label{sol3}
\end{equation}%
corresponding to the Hamiltonians (\ref{in2}) and (\ref{in7}), respectively,
(we give details of the proof in the Appendix~B), can be found by the method
proposed in \cite{Cor-Sot:Lop:Sua:Sus} in the form%
\begin{equation}
\psi =G\left( x,y,t\right) =\frac{1}{\sqrt{2\pi i\mu \left( t\right) }}\
e^{i\left( \alpha \left( t\right) x^{2}+\beta \left( t\right) xy+\gamma
\left( t\right) y^{2}\right) },  \label{sol4}
\end{equation}%
where%
\begin{eqnarray}
&&\alpha \left( t\right) =\frac{1}{4a\left( t\right) }\frac{\mu ^{\prime
}\left( t\right) }{\mu \left( t\right) }-\frac{d\left( t\right) }{2a\left(
t\right) },  \label{sol5} \\
&&\beta \left( t\right) =-\frac{1}{\mu \left( t\right) },\qquad \frac{%
d\gamma }{dt}+\frac{a\left( t\right) }{\mu \left( t\right) ^{2}}=0,
\label{sol6} \\
&&\gamma \left( t\right) =\frac{a\left( t\right) }{\mu \left( t\right) \mu
^{\prime }\left( t\right) }+\frac{d\left( 0\right) }{2a\left( 0\right) }%
-4\int_{0}^{t}\frac{a\left( \tau \right) \sigma \left( \tau \right) }{\left(
\mu ^{\prime }\left( \tau \right) \right) ^{2}}\ d\tau ,  \label{sol7}
\end{eqnarray}%
and the function $\mu \left( t\right) $ satisfies the\ characteristic
equation%
\begin{equation}
\mu ^{\prime \prime }-\tau \left( t\right) \mu ^{\prime }+4\sigma \left(
t\right) \mu =0  \label{sol8}
\end{equation}%
with%
\begin{equation}
\tau \left( t\right) =\frac{a^{\prime }}{a}-2c+4d,\qquad \sigma \left(
t\right) =ab-cd+d^{2}+\frac{d}{2}\left( \frac{a^{\prime }}{a}-\frac{%
d^{\prime }}{d}\right)  \label{sol9}
\end{equation}%
subject to the initial data%
\begin{equation}
\mu \left( 0\right) =0,\qquad \mu ^{\prime }\left( 0\right) =2a\left(
0\right) \neq 0.  \label{sol10}
\end{equation}%
Equation (\ref{sol5}) (more details can be found in \cite%
{Cor-Sot:Lop:Sua:Sus}) allows us to integrate the familiar Riccati nonlinear
differential equation emerging when one substitutes (\ref{sol4}) into (\ref%
{sol1}). See, for example, \cite{Haah:Stein}, \cite{Molch}, \cite{Rainville}%
, \cite{Rajah:Mah}, \cite{Wa} and references therein. A Hamiltonian
structure of these characteristic equations is discussed in Appendix~C.

In the case (\ref{sol2}), the characteristic equation has a special form of
Ince's equation \cite{Mag:Win}%
\begin{equation}
\mu ^{\prime \prime }+2\tan t\ \mu ^{\prime }-2\mu =0.  \label{sol11}
\end{equation}%
Two linearly independent solutions are found in \cite{Cor-Sot:Lop:Sua:Sus}:
\begin{eqnarray}
\mu _{1} &=&\cos t\cosh t+\sin t\sinh t=W\left( \cos t,\sinh t\right) ,
\label{sol12} \\
\mu _{2} &=&\cos t\sinh t+\sin t\cosh t=W\left( \cos t,\cosh t\right)
\label{sol13}
\end{eqnarray}%
with the Wronskian $W\left( \mu _{1},\mu _{2}\right) =2\cos ^{2}t=2a.$
Another method of integration of all characteristic equations from this
section is discussed in the Appendix~A; see Table~1 at the end of the paper
for the sets of fundamental solutions. The second case (\ref{sol3}) gives%
\begin{equation}
\mu ^{\prime \prime }-2\tanh t\ \mu ^{\prime }+2\mu =0  \label{sol14}
\end{equation}%
and the two linearly independent solutions are \cite{Sua:Sus:Vega}%
\begin{eqnarray}
\mu _{2} &=&\cos t\sinh t+\sin t\cosh t=W\left( \cos t,\cosh t\right) ,
\label{sol15} \\
\mu _{3} &=&\sin t\sinh t-\cos t\cosh t=W\left( \sin t,\cosh t\right)
\label{sol16}
\end{eqnarray}%
with $W\left( \mu _{2},\mu _{3}\right) =2\cosh ^{2}t=2a.$ Equation (\ref%
{sol14}) can be obtain from (\ref{sol11}) as a result of the substitution $%
t\rightarrow it.$ Also, $W\left( \mu _{1},\mu _{3}\right) =\sin \left(
2t\right) +\sinh \left( 2t\right) .$ The common solution of the both
characteristic equations, namely,%
\begin{equation}
\mu \left( t\right) =\mu _{2}=\cos t\sinh t+\sin t\cosh t,  \label{sol17}
\end{equation}%
satisfies the initial conditions (\ref{sol10}).

From (\ref{sol4})--(\ref{sol7}), as a result of elementary calculations, one
arrives at the Green function (\ref{in5}) in the case (\ref{sol2}) and has
to interchange there $x\leftrightarrow y$ in the second case (\ref{sol3}).
The reader can see some calculation details in section~9, where more general
solutions are found in a similar way. The next section explains this unusual
symmetry between two propagators from a more general point of view.

Two more completely integrable cases of the dual quadratic Hamiltonians
occur when%
\begin{equation}
a=\sin ^{2}t,\quad b=\cos ^{2}t,\quad c=2d=-\sin \left( 2t\right)
\label{sol18}
\end{equation}%
and%
\begin{equation}
a=\sinh ^{2}t,\quad b=\cosh ^{2}t,\quad c=2d=\sinh \left( 2t\right) .
\label{sol19}
\end{equation}%
The corresponding characteristic equations are%
\begin{equation}
\mu ^{\prime \prime }-2\cot t\ \mu ^{\prime }-2\mu =0  \label{sol20}
\end{equation}%
and%
\begin{equation}
\mu ^{\prime \prime }-2\coth t\ \mu ^{\prime }+2\mu =0,  \label{sol21}
\end{equation}%
respectively, with a common solution%
\begin{equation}
\mu \left( t\right) =\mu _{4}=\sin t\cosh t-\cos t\sinh t=W\left( \sin
t,\sinh t\right)  \label{sol22}
\end{equation}%
such that $\mu \left( 0\right) =\mu ^{\prime }\left( 0\right) =\mu ^{\prime
\prime }\left( 0\right) =0$ and $\mu ^{\prime \prime \prime }\left( 0\right)
=4.$ Once again, from (\ref{sol4})--(\ref{sol7}) one arrives at the Green
function%
\begin{eqnarray}
G\left( x,y,t\right) &=&\frac{1}{\sqrt{2\pi i\left( \sin t\cosh t-\cos
t\sinh t\right) }}  \label{sol23} \\
&&\times \exp \left( \frac{\left( x^{2}+y^{2}\right) \cos t\cosh
t-2xy+\left( x^{2}-y^{2}\right) \sin t\sinh t}{2i\left( \cos t\sinh t-\sin
t\cosh t\right) }\right)  \notag
\end{eqnarray}%
in the case (\ref{sol18}) and has to interchange there $x\leftrightarrow y$
in the second case (\ref{sol19}). The corresponding asymptotic formula takes
the form%
\begin{equation}
G\left( x,y,t\right) =\frac{e^{i\left( \alpha \left( t\right) x^{2}+\beta
\left( t\right) xy+\gamma \left( t\right) y^{2}\right) }}{\sqrt{2\pi i\mu
\left( t\right) }}\sim \frac{1}{\sqrt{4\pi i\varepsilon }}\exp \left( i%
\dfrac{\left( x-y\right) ^{2}}{4\varepsilon }\right)  \label{sol24}
\end{equation}%
as $\varepsilon =t^{3}/3\rightarrow 0^{+}.$ We will show in section~6, see
Eqs.~(\ref{mr1}) and (\ref{mr6}), that our cases (\ref{sol2})--(\ref{sol3})
and (\ref{sol18})--(\ref{sol19}) are related to each other by means of the
Fourier transform.

We have considered some elementary solutions of the characteristic equation (%
\ref{sol8}), which are of interest in this paper. Generalizations to the
forced modified oscillators are obvious; see Ref.~\cite{Me:Co:Su}. More
complicated cases may include special functions, like Bessel, hypergeometric
or elliptic functions \cite{An:As:Ro}, \cite{Cor-Sot:Lop:Sua:Sus}, \cite%
{Ni:Uv}, \cite{Rain}, \cite{Sus:Trey}, and \cite{Wa}.

\section{On A \textquotedblleft Hidden\textquotedblright\ Symmetry of
Quadratic Propagators}

Here we shall elaborate on the symmetry of propagators with respect to the
substitution $x\leftrightarrow y.$

\begin{lemma}
Consider two time-dependent Schr\"{o}dinger equations with quadratic
Hamiltonians%
\begin{equation}
i\frac{\partial \psi }{\partial t}=-a_{k}\left( t\right) \frac{\partial
^{2}\psi }{\partial x^{2}}+b_{k}\left( t\right) x^{2}\psi -i\left(
c_{k}\left( t\right) x\frac{\partial \psi }{\partial x}+d_{k}\left( t\right)
\psi \right) \quad (k=1,2),  \label{sym1}
\end{equation}%
where $c_{1}-2d_{1}=c_{2}-2d_{2}=\varepsilon \left( t\right) $ and $%
d_{k}\left( 0\right) =0.$ Suppose that the initial value problems for
corresponding characteristic equations%
\begin{equation}
\mu ^{\prime \prime }-\tau _{k}\left( t\right) \mu ^{\prime }+4\sigma
_{k}\left( t\right) \mu =0,\qquad \mu \left( 0\right) =0,\quad \mu ^{\prime
}\left( 0\right) =2a_{k}\left( 0\right) \neq 0  \label{sym2}
\end{equation}%
with%
\begin{equation}
\tau _{k}\left( t\right) =\frac{a_{k}^{\prime }}{a_{k}}-2c_{k}+4d_{k},\quad
\sigma _{k}\left( t\right) =a_{k}b_{k}-c_{k}d_{k}+d_{k}^{2}+\frac{d_{k}}{2}%
\left( \frac{a_{k}^{\prime }}{a_{k}}-\frac{d_{k}^{\prime }}{d_{k}}\right)
\label{sym3}
\end{equation}%
have a joint solution $\mu \left( t\right) $ and, in addition, the following
relations hold%
\begin{equation}
4\left( a_{1}b_{1}-c_{1}d_{1}+d_{1}^{2}\right) =\frac{4a_{1}a_{2}h^{2}-%
\left( \mu ^{\prime }\right) ^{2}}{\mu ^{2}}-2\varepsilon \frac{\mu ^{\prime
}}{\mu }=4\left( a_{2}b_{2}-c_{2}d_{2}+d_{2}^{2}\right) ,  \label{sym4}
\end{equation}%
where $h\left( t\right) =\exp \left( -\dint\nolimits_{\!0}^{t}\varepsilon
\left( \tau \right) \ d\tau \right) .$ Then the corresponding fundamental
solutions%
\begin{equation}
\psi _{k}=G_{k}\left( x,y,t\right) =\frac{1}{\sqrt{2\pi i\mu \left( t\right)
}}\ e^{i\left( \alpha _{k}\left( t\right) x^{2}+\beta _{k}\left( t\right)
xy+\gamma _{k}\left( t\right) y^{2}\right) }  \label{sym5}
\end{equation}%
possess the following symmetry%
\begin{equation}
\alpha \left( t\right) =\alpha _{1}\left( t\right) =\gamma _{2}\left(
t\right) ,\quad \gamma \left( t\right) =\gamma _{1}\left( t\right) =\alpha
_{2}\left( t\right) ,\quad \beta \left( t\right) =\beta _{1}\left( t\right)
=\beta _{2}\left( t\right)  \label{sym6}
\end{equation}%
and%
\begin{equation}
G_{1}\left( x,y,t\right) =G_{2}\left( y,x,t\right) .  \label{sym7}
\end{equation}%
This property holds for a single Schr\"{o}dinger equation under the single
hypothesis (\ref{sym4}).
\end{lemma}

Indeed, according to Ref.~\cite{Cor-Sot:Lop:Sua:Sus},%
\begin{equation}
\beta \left( t\right) =\beta _{1}\left( t\right) =\beta _{2}\left( t\right)
=-\frac{h\left( t\right) }{\mu \left( t\right) }  \label{sym8}
\end{equation}%
in the case of a joint solution $\mu \left( t\right) $ of two characteristic
equations. In view of the structure of propagators for general quadratic
Hamiltonians found in \cite{Cor-Sot:Lop:Sua:Sus}, the symmetry under
consideration holds if we have%
\begin{equation}
\alpha =\frac{1}{4a_{1}}\frac{\mu ^{\prime }}{\mu }-\frac{d_{1}}{2a_{1}}%
,\qquad \frac{d\alpha }{dt}+a_{2}\frac{h^{2}}{\mu ^{2}}=0  \label{sym9}
\end{equation}%
and%
\begin{equation}
\gamma =\frac{1}{4a_{2}}\frac{\mu ^{\prime }}{\mu }-\frac{d_{2}}{2a_{2}}%
,\qquad \frac{d\gamma }{dt}+a_{1}\frac{h^{2}}{\mu ^{2}}=0,  \label{sym10}
\end{equation}%
simultaneously. Excluding $\alpha $ from (\ref{sym9}), one gets%
\begin{equation}
\mu ^{\prime \prime }-\frac{a_{1}^{\prime }}{a_{1}}\mu ^{\prime
}+2d_{1}\left( \frac{a_{1}^{\prime }}{a_{1}}-\frac{d_{1}^{\prime }}{d_{1}}%
\right) \mu =\frac{\left( \mu ^{\prime }\right) ^{2}-4a_{1}a_{2}h^{2}}{\mu }.
\label{sym11}
\end{equation}%
Comparison with the characteristic equation results in the first condition
in (\ref{sym4}). The case of $\gamma ,$ which gives the second condition, is
similar. This completes the proof.

A few examples are in order. When $a=1/2,$ $b=c=d=0,$ and $\mu ^{\prime
\prime }=0,$ $\mu =t,$ one gets%
\begin{equation}
G\left( x,y,t\right) =\frac{1}{\sqrt{2\pi it}}\ \exp \left( \frac{i\left(
x-y\right) ^{2}}{2t}\right)  \label{sym12}
\end{equation}%
as the free particle propagator \cite{Fey:Hib} with an obvious symmetry
under consideration. Our criteria (\ref{sym4}), namely, $4a^{2}=\left( \mu
^{\prime }\right) ^{2},$ stands.

The simple harmonic oscillator with $a=b=1/2,$ $c=d=0$ and $\mu ^{\prime
\prime }+\mu =0,$ $\mu =\sin t$ has the familiar propagator of the form%
\begin{equation}
G\left( x,y,t\right) =\frac{1}{\sqrt{2\pi i\sin t}}\ \exp \left( \dfrac{i}{%
2\sin t}\left( \left( x^{2}+y^{2}\right) \cos t-2xy\right) \right) ,
\label{sym13}
\end{equation}%
which is studied in detail at \cite{Beauregard}, \cite{Gottf:T-MY}, \cite%
{Holstein}, \cite{Maslov:Fedoriuk}, \cite{Merz}, \cite{Thomber:Taylor}. (For
an extension to the case of the forced harmonic oscillator including an
extra velocity-dependent term and a time-dependent frequency, see \cite%
{FeynmanPhD}, \cite{Feynman}, \cite{Fey:Hib} and \cite{Lop:Sus}.) Our
condition (\ref{sym4}) takes the form of the trigonometric identity%
\begin{equation}
4ab=\frac{4a^{2}-\left( \mu ^{\prime }\right) ^{2}}{\mu ^{2}},  \label{sym14}
\end{equation}%
which confirms the symmetry of the propagator.

For the quantum damped oscillator \cite{Cor-Sot:Sua:Sus} $a=b=\omega _{0}/2,$
$c=d=-\lambda $ and%
\begin{eqnarray}
G\left( x,y,t\right) &=&\sqrt{\frac{\omega e^{\lambda t}}{2\pi i\omega
_{0}\sin \omega t}}\ \exp \left( \dfrac{i\omega }{2\omega _{0}\sin \omega t}%
\left( \left( x^{2}+y^{2}\right) \cos \omega t-2xy\right) \right)  \notag
\label{sym14a} \\
&&\times \exp \left( \frac{i\lambda }{2\omega _{0}}\left( x^{2}-y^{2}\right)
\right)  \label{sym14ab}
\end{eqnarray}%
with $\omega =\sqrt{\omega _{0}^{2}-\lambda ^{2}}>0$ and $\mu =\left( \omega
_{0}/\omega \right) e^{-\lambda t}\sin \omega t.$ The criterion%
\begin{equation}
4ab=\frac{4\left( ah\right) ^{2}-\left( \mu ^{\prime }\right) ^{2}}{\mu ^{2}}%
-2\varepsilon \frac{\mu ^{\prime }}{\mu },  \label{sym14b}
\end{equation}%
where $\varepsilon =c-2d=\lambda $ and $h=$ $e^{-\lambda t},$ holds. But
here $d\left( 0\right) =-\lambda \neq 0$ and a more detailed analysis of
asymptotics gives an extra antisymmetric term in the propagator above; see
\cite{Cor-Sot:Sua:Sus} for more details.

The case of Hamiltonians (\ref{in2}) and (\ref{in7}) corresponds to%
\begin{equation}
a_{1}=\cos ^{2}t,\quad b_{1}=\sin ^{2}t,\quad c_{1}=2d_{1}=\sin \left(
2t\right)  \label{sym15}
\end{equation}%
and%
\begin{equation}
a_{2}=\cosh ^{2}t,\quad b_{2}=\sinh ^{2}t,\quad c_{2}=2d_{2}=-\sinh \left(
2t\right) ,  \label{sym16}
\end{equation}%
respectively. Our criteria (\ref{sym4}) are satisfied in view of an obvious
identity%
\begin{equation}
4a_{1}a_{2}=\left( \mu ^{\prime }\right) ^{2}.  \label{sym17}
\end{equation}%
The characteristic function is given by (\ref{sol17}). This explains the
propagator symmetry found in the previous section.

Our last dual pair of quadratic Hamiltonians has the following coefficients%
\begin{equation}
a_{1}=\sin ^{2}t,\quad b_{1}=\cos ^{2}t,\quad c_{1}=2d_{1}=-\sin \left(
2t\right)  \label{sym17a}
\end{equation}%
and%
\begin{equation}
a_{2}=\sinh ^{2}t,\quad b_{2}=\cosh ^{2}t,\quad c_{2}=2d_{2}=\sinh \left(
2t\right) .  \label{sym17b}
\end{equation}%
The criteria (\ref{sym4}) are satisfied in view of the identity (\ref{sym17}%
) with the characteristic function (\ref{sol22}) and, therefore, the
propagator (\ref{sol23}) obeys the symmetry under the substitution $%
x\leftrightarrow y.$

\begin{remark}
A simple relation%
\begin{equation}
\frac{\mu ^{\prime }}{\mu }=4\frac{\sigma _{1}-\sigma _{2}}{\tau _{1}-\tau
_{2}},  \label{sym18}
\end{equation}%
which is valid for a joint solution of two characteristic equations, can be
used in our criteria (\ref{sym4}).
\end{remark}

Although we have formulated the hypotheses of our lemma for the Green
functions only, it can be applied to solutions with regular initial data.
For instance, a pair of characteristic equations (\ref{sol11}) and (\ref%
{sol21}) has a joint solution given by (\ref{sol12}), which does not satisfy
initial conditions required for the Green functions. The coefficients of the
corresponding quadratic Hamiltonians are%
\begin{equation}
a_{1}=\cos ^{2}t,\quad b_{1}=\sin ^{2}t,\quad c_{1}=2d_{1}=\sin \left(
2t\right)  \label{sym19}
\end{equation}%
and%
\begin{equation}
a_{2}=\sinh ^{2}t,\quad b_{2}=\cosh ^{2}t,\quad c_{2}=2d_{2}=\sinh \left(
2t\right) .  \label{sym20}
\end{equation}%
The criteria (\ref{sym4}) are satisfied in view of the identity (\ref{sym17}%
) and the particular solution%
\begin{eqnarray}
&&\psi =K\left( x,y,t\right) =\frac{1}{\sqrt{2\pi \left( \cos t\cosh t+\sin
t\sinh t\right) }}  \label{sym21} \\
&&\qquad \times \exp \left( \frac{\left( x^{2}+y^{2}\right) \sin t\cosh
t-2xy-\left( x^{2}-y^{2}\right) \cos t\sinh t}{2i\left( \cos t\cosh t+\sin
t\sinh t\right) }\right)  \notag
\end{eqnarray}%
obeys the symmetry under the substitution $x\leftrightarrow y.$ The initial
condition is the standing wave $K\left( x,y,0\right) =e^{ixy}/\sqrt{2\pi }.$

In a similar fashion, the characteristic equations (\ref{sol20}) and (\ref%
{sol14}) have a common solution%
\begin{equation}
\mu =-\mu _{3}=\cos t\cosh t-\sin t\sinh t.  \label{sym22}
\end{equation}%
The coefficients of the corresponding Hamiltonians are%
\begin{equation}
a_{1}=\sin ^{2}t,\quad b_{1}=\cos ^{2}t,\quad c_{1}=2d_{1}=-\sin \left(
2t\right)  \label{sym23}
\end{equation}%
and%
\begin{equation}
a_{2}=\cosh ^{2}t,\quad b_{2}=\sinh ^{2}t,\quad c_{2}=2d_{2}=-\sinh \left(
2t\right) .  \label{sym24}
\end{equation}%
The criteria (\ref{sym4}) are satisfied once again and the particular
solution is given by%
\begin{eqnarray}
&&\psi =K\left( x,y,t\right) =\frac{1}{\sqrt{2\pi \left( \cos t\cosh t-\sin
t\sinh t\right) }}  \label{sym25} \\
&&\qquad \times \exp \left( \frac{\left( x^{2}+y^{2}\right) \sin t\cosh
t+2xy+\left( x^{2}-y^{2}\right) \cos t\sinh t}{2i\left( \cos t\cosh t-\sin
t\sinh t\right) }\right)  \notag
\end{eqnarray}%
with $K\left( x,y,0\right) =e^{-ixy}/\sqrt{2\pi }.$ We shall discuss in
section~6 how these solutions are related to the corresponding time
evolution operators.

\section{Complex Form of The Propagators}

It is worth noting that the propagator in (\ref{in5}) can be rewritten in
terms of the Wronskians of trigonometric and hypergeometric functions as%
\begin{eqnarray}
G\left( x,y,t\right) &=&\frac{1}{\sqrt{2\pi iW\left( \cos t,\cosh t\right) }}
\label{cf1} \\
&&\times \exp \left( \frac{W\left( \sin t,\cosh t\right) x^{2}+2xy-W\left(
\cos t,\sinh t\right) y^{2}}{2iW\left( \cos t,\cosh t\right) }\right) .
\notag
\end{eqnarray}%
This simply means that our propagator has the following structure%
\begin{equation}
G=\sqrt{\frac{c_{2}-c_{3}}{4\pi i}}\ \exp \left( \frac{c_{1}x^{2}+\left(
c_{2}-c_{3}\right) xy-c_{4}y^{2}}{2i}\right) ,  \label{cf2}
\end{equation}%
where the coefficients are solutions of the system of linear equations%
\begin{eqnarray}
c_{1}\cos t+c_{2}\cosh t &=&\sin t,  \label{cf3} \\
-c_{1}\sin t+c_{2}\sinh t &=&\cos t,  \notag \\
c_{3}\cos t+c_{4}\cosh t &=&\sinh t,  \notag \\
-c_{3}\sin t+c_{4}\sinh t &=&\cosh t  \notag
\end{eqnarray}%
obtained by Cramer's rule. A complex form of this system is%
\begin{equation}
c_{1}z^{\ast }+c_{2}\zeta =iz^{\ast },\quad c_{3}z^{\ast }+c_{4}\zeta
=i\zeta ^{\ast },  \label{cf4}
\end{equation}%
where we introduce two complex variables%
\begin{equation}
z=\cos t+i\sin t,\qquad \zeta =\cosh t+i\sinh t  \label{cf5}
\end{equation}%
and use the star for complex conjugate. Taking the complex conjugate of the
system (\ref{cf4}), which has the real-valued solutions, and using Cramer's
rule once again, one gets%
\begin{eqnarray}
&&c_{1}=\frac{z\zeta +z^{\ast }\zeta ^{\ast }}{i\left( z\zeta -z^{\ast
}\zeta ^{\ast }\right) },\qquad c_{2}=\frac{2i}{z\zeta -z^{\ast }\zeta
^{\ast }},  \label{cf6} \\
&&c_{3}=\frac{2}{i\left( z\zeta -z^{\ast }\zeta ^{\ast }\right) },\qquad
c_{4}=-\frac{z\zeta ^{\ast }+z^{\ast }\zeta }{i\left( z\zeta -z^{\ast }\zeta
^{\ast }\right) }.  \notag
\end{eqnarray}%
\noindent As a result, we obtain a compact symmetric expression of the
propagator (\ref{in5}) as a function of two complex variables%
\begin{eqnarray}
G\left( x,y,t\right) &=&G\left( x,y,z,\zeta \right) =\frac{1}{\sqrt{\pi
\left( z\zeta -z^{\ast }\zeta ^{\ast }\right) }}  \label{cf7} \\
&&\times \exp \left( \frac{\left( z\zeta +z^{\ast }\zeta ^{\ast }\right)
x^{2}-4xy+\left( z\zeta ^{\ast }+z^{\ast }\zeta \right) y^{2}}{2\left(
z^{\ast }\zeta ^{\ast }-z\zeta \right) }\right) .  \notag
\end{eqnarray}%
%
%
\begin{figure}[ptbh]
\centering\scalebox{.75}{\includegraphics{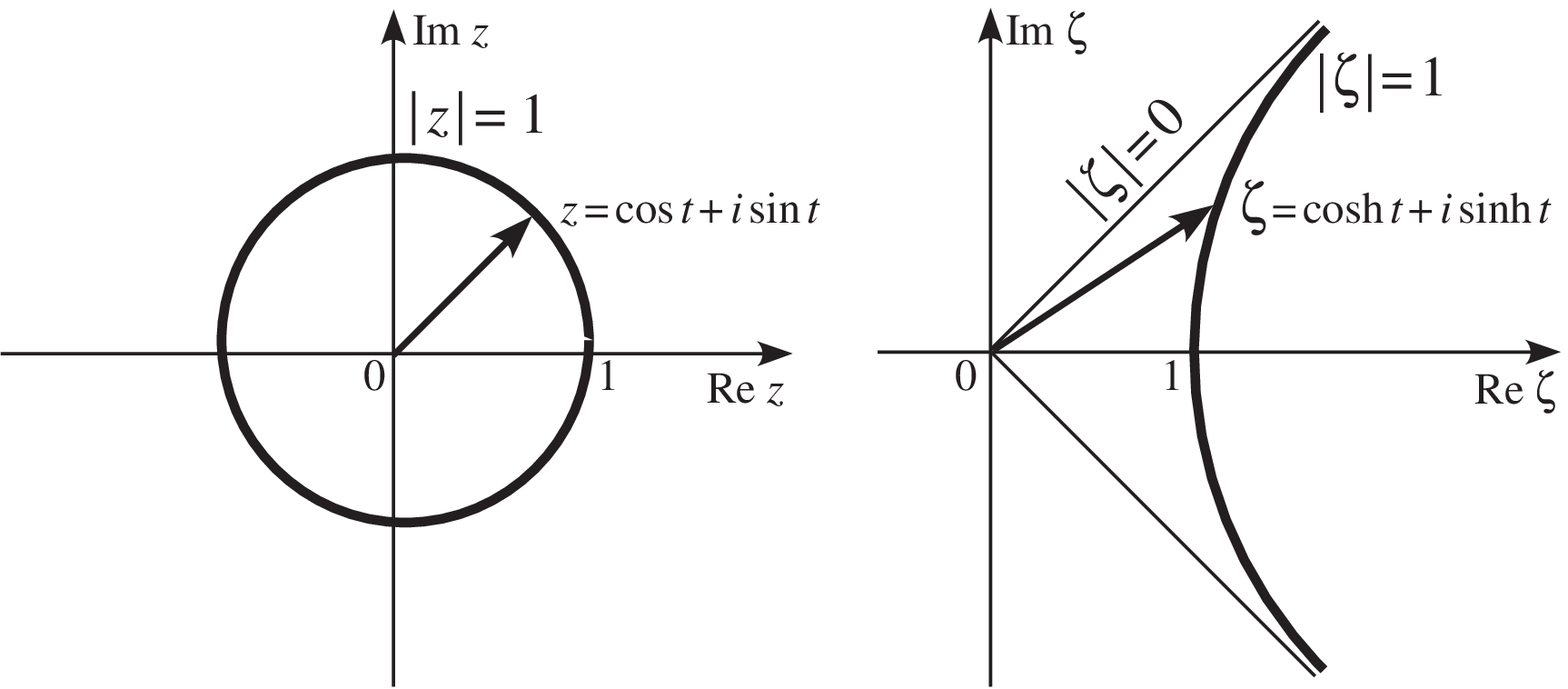}}
\caption{Two synchronized \textquotedblleft clocks\textquotedblright , or
contours in complex Euclidean $z$ and pseudo-Euclidean $\protect\zeta $
\textquotedblleft time\textquotedblright\ planes, corresponding to modified
oscillators.}
\end{figure}
%
This function takes a familiar form%
\begin{equation}
G=\frac{1}{\sqrt{2\pi i\left( x_{1}x_{4}+x_{2}x_{3}\right) }}\exp \left(
\frac{\left( x^{2}-y^{2}\right) x_{2}x_{4}+2xy-\left( x^{2}+y^{2}\right)
x_{1}x_{3}}{2i\left( x_{1}x_{4}+x_{2}x_{3}\right) }\right) ,  \label{cf8}
\end{equation}%
in a real-valued four-dimensional vector space, if we set $z=x_{1}+ix_{2}$
and $\zeta =x_{3}+ix_{4}$ with $x_{1}^{\prime }=-x_{2},$ $x_{2}^{\prime
}=x_{1},$ $x_{3}^{\prime }=x_{4},$ $x_{4}^{\prime }=x_{3},$ and solve the
following initial value problem %
%
\begin{eqnarray}
&&x_{1}^{\prime \prime }+x_{1}=0,\qquad x_{1}\left( 0\right) =1,\quad
x_{1}^{\prime }\left( 0\right) =0,  \label{cf9} \\
&&x_{2}^{\prime \prime }+x_{2}=0,\qquad x_{2}\left( 0\right) =0,\quad
x_{2}^{\prime }\left( 0\right) =1,  \notag \\
&&x_{3}^{\prime \prime }-x_{3}=0,\qquad x_{3}\left( 0\right) =1,\quad
x_{3}^{\prime }\left( 0\right) =0,  \notag \\
&&x_{4}^{\prime \prime }-x_{4}=0,\qquad x_{4}\left( 0\right) =0,\quad
x_{4}^{\prime }\left( 0\right) =1,  \notag
\end{eqnarray}%
the solution of which can be interpreted as a trajectory of a classical
\textquotedblleft particle\textquotedblright\ moving in this space; cf.~(\ref%
{in5}).

It is worth noting that our propagators expression (\ref{cf7}), extended to
a function of two independent complex variables $z$ and $\zeta ,$ allows us
to unify several exactly solvable quantum mechanical models in geometrical
terms, namely, by choosing different contours, with certain
\textquotedblleft synchronized\textquotedblright\ parametrization, in the
pair of complex \textquotedblleft time\textquotedblright\ planes under
consideration. Indeed, the free particle propagator (\ref{sym12}) appears in
this way when one chooses $z=1$ and $\zeta =1+it.$ The simple harmonic
oscillator propagator (\ref{sym13}) corresponds to the case $z=1$ and $\zeta
=e^{it}.$ As we have seen in this section, the propagator (\ref{in5}) is
also a special case of (\ref{cf7}). This is why we may refer to the
Hamiltonians under consideration as the ones of modified oscillators. By a
vague analogy with the special theory of relativity, one may also say that
in this case there are two synchronized \textquotedblleft
clocks\textquotedblright , namely, the two contour parameterized by (\ref%
{cf5}), one in Euclidean and another one in the pseudo-Euclidean two
dimensional spaces, respectively, which geometrically describes a time
evolution for the Hamiltonians of modified oscillators; see Figure~1. This
idea of introducing a geometric structure of time in the problem under
consideration may be useful for other types of evolutionary equations. %
%
%
\begin{figure}[ptbh]
\centering\scalebox{.75}{\includegraphics{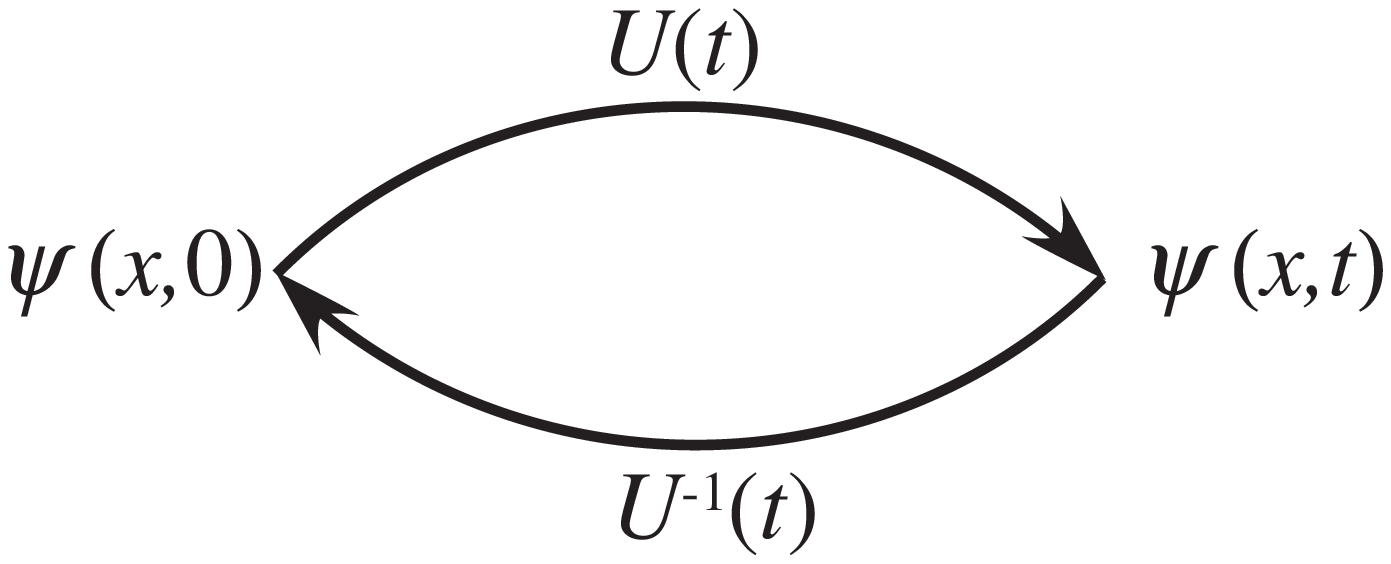}}
\caption{The time evolution operator and its inverse.}
\end{figure}
%

In a similar fashion, our new propagator (\ref{sol23}) can be rewritten in
terms of the Wronskians as%
\begin{eqnarray}
G\left( x,y,t\right) &=&\frac{1}{\sqrt{2\pi iW\left( \sin t,\sinh t\right) }}
\label{cf10} \\
&&\times \exp \left( \frac{W\left( \cos t,\sinh t\right) x^{2}-2xy-W\left(
\sin t,\cosh t\right) y^{2}}{2iW\left( \sinh t,\sin t\right) }\right) .
\notag
\end{eqnarray}%
The corresponding complex form is%
\begin{eqnarray}
G\left( x,y,t\right) &=&G\left( x,y,z,\zeta \right) =\frac{1}{\sqrt{\pi
\left( z\zeta ^{\ast }-z^{\ast }\zeta \right) }} \\
&&\times \exp \left( \frac{\left( z\zeta ^{\ast }+z^{\ast }\zeta \right)
x^{2}-4xy+\left( z\zeta +z^{\ast }\zeta ^{\ast }\right) y^{2}}{2\left(
z^{\ast }\zeta -z\zeta ^{\ast }\right) }\right) .  \notag
\end{eqnarray}%
We leave the details to the reader.

\section{The Inverse Operator and Time Reversal}

We follow the method suggested in \cite{Sua:Sus} for general quadratic
Hamiltonians with somewhat different details. The left inverse of the
integral operator defined by (\ref{in4})--(\ref{in5}), namely,%
\begin{equation}
U\left( t\right) \psi \left( x\right) =\int_{-\infty }^{\infty }G\left(
x,y,t\right) \ \psi \left( y\right) \ dy,  \label{inv1}
\end{equation}%
is%
\begin{equation}
U^{-1}\left( t\right) \chi \left( x\right) =\int_{-\infty }^{\infty }G\left(
y,x,-t\right) \ \chi \left( y\right) \ dy  \label{inv2}
\end{equation}%
in view of $U^{-1}=U^{\dagger };$ see Figure~2. Indeed, when $s<t,$ by the
Fubini theorem%
\begin{eqnarray}
U^{-1}\left( s\right) \left( U\left( t\right) \psi \right) &=&U^{-1}\left(
s\right) \chi =\int_{-\infty }^{\infty }G\left( z,x,-s\right) \ \chi \left(
z\right) \ dz  \label{inv3} \\
&=&\int_{-\infty }^{\infty }G\left( z,x,-s\right) \ \left( \int_{-\infty
}^{\infty }G\left( z,y,t\right) \ \psi \left( y\right) \ dy\right) \ dz
\notag \\
&=&\int_{-\infty }^{\infty }G\left( x,y,s,t\right) \ \psi \left( y\right) \
dy.  \notag
\end{eqnarray}%
\noindent Here%
\begin{eqnarray}
G\left( x,y,s,t\right) &=&\int_{-\infty }^{\infty }G\left( z,x,-s\right)
G\left( z,y,t\right) \ dz  \label{inv4} \\
&=&\frac{e^{i\left( \gamma \left( t\right) y^{2}-\gamma \left( s\right)
x^{2}\right) }}{2\pi \sqrt{\mu \left( s\right) \mu \left( t\right) }}%
\int_{-\infty }^{\infty }e^{i\left( \left( \alpha \left( t\right) -\alpha
\left( s\right) \right) z^{2}+\left( \beta \left( t\right) y-\beta \left(
s\right) x\right) z\right) }\ dz  \notag \\
&=&\frac{1}{\sqrt{4\pi i\mu \left( s\right) \mu \left( t\right) \left(
\alpha \left( s\right) -\alpha \left( t\right) \right) }}  \notag \\
&&\times \exp \left( \frac{\left( \beta \left( t\right) y-\beta \left(
s\right) x\right) ^{2}-4\left( \alpha \left( t\right) -\alpha \left(
s\right) \right) \left( \gamma \left( t\right) y^{2}-\gamma \left( s\right)
x^{2}\right) }{4i\left( \alpha \left( t\right) -\alpha \left( s\right)
\right) }\right)  \notag
\end{eqnarray}%
by the familiar Gaussian integral \cite{Bo:Shi}, \cite{Palio:Mead} and \cite%
{Ru}:%
\begin{equation}
\int_{-\infty }^{\infty }e^{i\left( az^{2}+2bz\right) }\,dz=\sqrt{\frac{\pi i%
}{a}}\,e^{-ib^{2}/a}.  \label{gauss}
\end{equation}%
In view of (\ref{sol6}),%
\begin{equation}
\left( \beta \left( t\right) y-\beta \left( s\right) x\right) ^{2}=\left(
\frac{x}{\mu \left( s\right) }-\frac{y}{\mu \left( t\right) }\right) ^{2}=%
\frac{\left( \sqrt{\dfrac{\mu \left( t\right) }{\mu \left( s\right) }}x-%
\sqrt{\dfrac{\mu \left( s\right) }{\mu \left( t\right) }}y\right) ^{2}}{\mu
\left( s\right) \mu \left( t\right) \left( \alpha \left( t\right) -\alpha
\left( s\right) \right) },  \label{inv5}
\end{equation}%
and a singular part of (\ref{inv4}) becomes%
\begin{eqnarray*}
&&\frac{1}{\sqrt{4\pi i\mu \left( s\right) \mu \left( t\right) \left( \alpha
\left( s\right) -\alpha \left( t\right) \right) }}\exp \left( \frac{\left(
\beta \left( t\right) y-\beta \left( s\right) x\right) ^{2}}{4i\left( \alpha
\left( t\right) -\alpha \left( s\right) \right) }\right) \\
&&\quad =\frac{1}{\sqrt{4\pi i\mu \left( s\right) \mu \left( t\right) \left(
\alpha \left( s\right) -\alpha \left( t\right) \right) }}\exp \left( \frac{%
\left( \sqrt{\frac{\mu \left( t\right) }{\mu \left( s\right) }}x-\sqrt{\frac{%
\mu \left( s\right) }{\mu \left( t\right) }}y\right) ^{2}}{4i\mu \left(
s\right) \mu \left( t\right) \left( \alpha \left( t\right) -\alpha \left(
s\right) \right) }\right) .
\end{eqnarray*}%
Thus, in the limit $s\rightarrow t^{-},$ one can obtain formally the
identity operator in the right hand side of (\ref{inv3}). The leave the
details to the reader.

On the other hand, the integral operator in (\ref{in4})--(\ref{in5}), namely,%
\begin{equation}
\chi \left( x\right) =\frac{1}{\sqrt{2\pi i\mu \left( t\right) }}%
\int_{-\infty }^{\infty }e^{i\left( \alpha \left( t\right) x^{2}+\beta
\left( t\right) xy+\gamma \left( t\right) y^{2}\right) }\ \psi \left(
y\right) \ dy  \label{fourier}
\end{equation}%
is essentially the Fourier transform and its inverse is given by%
\begin{equation}
\psi \left( y\right) =\frac{1}{\sqrt{-2\pi i\mu \left( t\right) }}%
\int_{-\infty }^{\infty }e^{-i\left( \alpha \left( t\right) x^{2}+\beta
\left( t\right) xy+\gamma \left( t\right) y^{2}\right) }\ \chi \left(
x\right) \ dy  \label{fourierinv}
\end{equation}%
in correspondence with our definition (\ref{inv2}) in view of (\ref{sol6}).

The Schr\"{o}dinger equation (\ref{in1}) retains the same form if we replace
$t$ in it by $-t$ and, at the same time, take complex conjugate provided
that $\left( H\left( -t\right) \varphi \right) ^{\ast }=H\left( t\right)
\varphi ^{\ast }.$ The last relation holds for both Hamiltonians (\ref{in2})
and (\ref{in7}). Hence the function $\chi \left( x,t\right) =\psi ^{\ast
}\left( x,-t\right) $ does satisfy the same equation as the original wave
function $\psi \left( x,t\right) .$ This property is usually known as the
symmetry with respect to time inversion (time reversal) in quantum mechanics
\cite{Gottf:T-MY}, \cite{La:Lif}, \cite{Merz}, \cite{Wigner}. This fact is
obvious from a general solution given by (\ref{in4})--(\ref{in5}) for our
Hamiltonians.

On the other hand, by the definition, the (left) inverse $U^{-1}\left(
t\right) $ of the time evolution operator $U\left( t\right) $ returns the
system to its initial quantum state:
\begin{eqnarray}
&&\psi \left( x,t\right) =U\left( t\right) \psi \left( x,0\right) ,
\label{inv6} \\
&&U^{-1}\left( t\right) \psi \left( x,t\right) =U^{-1}\left( t\right) \left(
U\left( t\right) \psi \left( x,0\right) \right) =\psi \left( x,0\right) .
\label{inv7}
\end{eqnarray}%
Our analysis of two oscillator models under consideration shows that this
may be related to the reversal of time in the following manner. The left
inverse of the time evolution operator (\ref{in4}) for the Schr\"{o}dinger
equation (\ref{in1}) with the original Hamiltonian of a modified oscillator (%
\ref{in2}) can be obtained by the time inversion $t\rightarrow -t$ in the
evolution operator corresponding to the new \textquotedblleft
dual\textquotedblright\ Hamiltonian (\ref{in7}) (and vise versa). The same
is true for the second pair of dual Hamiltonians. More details will be given
in section~6. This is an example of a situation in mathematical physics and
quantum mechanics when a change in the direction of time may require a total
change of the system dynamics in order to return the system back to its
original quantum state. Moreover, moving backward in time the system will
repeat the same quantum states only when%
\begin{equation}
\psi \left( x,t-s\right) =U\left( t-s\right) \psi \left( x,0\right)
=U^{-1}\left( s\right) U\left( t\right) \psi \left( x,0\right) ,\qquad 0\leq
s\leq t,  \label{inv8}
\end{equation}%
which is equivalent to the semi-group property%
\begin{equation}
U\left( s\right) U\left( t-s\right) =U\left( t\right)  \label{inv9}
\end{equation}%
for the time evolution operator. This seems not true for propagators (\ref%
{in5}) and (\ref{sol23}).

\section{The Momentum Representation}

The time-dependent Schr\"{o}dinger equation (\ref{sol1}) can be rewritten in
terms of the operator of coordinate $x$ and the operator of linear momentum $%
p_{x}=i^{-1}\partial /\partial x$ as follows%
\begin{equation}
i\frac{\partial \psi }{\partial t}=\left( a\left( t\right) p_{x}^{2}+b\left(
t\right) x^{2}+d\left( t\right) \left( xp_{x}+p_{x}x\right) \right) \psi
\label{mr1}
\end{equation}%
with $c=2d.$ The corresponding quadratic Hamiltonian%
\begin{equation}
H=a\left( t\right) p_{x}^{2}+b\left( t\right) x^{2}+d\left( t\right) \left(
xp_{x}+p_{x}x\right)  \label{mr2}
\end{equation}%
obeys a special symmetry, namely, it formally preserves this structure under
the permutation $x\leftrightarrow p_{x}.$ This fact is well-known for the
simple harmonic oscillator \cite{Gottf:T-MY}, \cite{La:Lif}, \cite{Merz}.

In order to interchange the coordinate and momentum operators in quantum
mechanics one switches between the coordinate and momentum representations
by means of the Fourier transform%
\begin{equation}
\psi \left( x\right) =\frac{1}{\sqrt{2\pi }}\int_{-\infty }^{\infty
}e^{ixy}\chi \left( y\right) \ dy=F\left[ \chi \right]  \label{mr3}
\end{equation}%
and its inverse%
\begin{equation}
\chi \left( y\right) =\frac{1}{\sqrt{2\pi }}\int_{-\infty }^{\infty
}e^{-ixy}\psi \left( x\right) \ dx=F^{-1}\left[ \psi \right] .  \label{mr4}
\end{equation}%
Indeed, the familiar properties%
\begin{equation}
p_{x}\psi =p_{x}F\left[ \chi \right] =F\left[ y\chi \right] ,\qquad x\psi =xF%
\left[ \chi \right] =-F\left[ p_{y}\chi \right]  \label{mr5}
\end{equation}%
imply%
\begin{equation}
p_{x}^{2}\psi =F\left[ y^{2}\chi \right] ,\qquad x^{2}\psi =F\left[
p_{y}^{2}\chi \right]  \label{mr5a}
\end{equation}%
and%
\begin{equation}
\left( xp_{x}+p_{x}x\right) \psi =-F\left[ \left( p_{y}y+yp_{y}\right) \chi %
\right] .  \label{mr5b}
\end{equation}%
Therefore,%
\begin{eqnarray*}
H\psi &=&\left( ap_{x}^{2}+bx^{2}+d\left( xp_{x}+p_{x}x\right) \right) F
\left[ \chi \right] \\
&=&F\left[ \left( bp_{y}^{2}+ay^{2}-d\left( yp_{y}+p_{y}y\right) \right)
\chi \right]
\end{eqnarray*}%
by the linearity of the Fourier transform. In view of%
\begin{equation*}
\frac{\partial \psi }{\partial t}=F\left[ \frac{\partial \chi }{\partial t}%
\right]
\end{equation*}%
the Schr\"{o}dinger equation (\ref{mr1}) takes the form%
\begin{equation}
i\frac{\partial \chi }{\partial t}=\left( b\left( t\right) p_{y}^{2}+a\left(
t\right) y^{2}-d\left( t\right) \left( yp_{y}+p_{y}y\right) \right) \chi
\label{mr6}
\end{equation}%
with $a\leftrightarrow b$ and $d\rightarrow -d$ in the momentum
representation.

This property finally reveals that our quadratic Hamiltonians (\ref{sol2})
and (\ref{sol18}), similarly (\ref{sol3}) and (\ref{sol19}), corresponds to
the same Schr\"{o}dinger equation written in the coordinate and momentum
representations, respectively. Thus, in section~2, we have solved the Cauchy
initial value problem for modified oscillators both in the coordinate and
momentum representations.

In this paper the creation and annihilation operators are defined by%
\begin{eqnarray}
&&a^{\dagger }=\frac{p_{x}+ix}{\sqrt{2}}=\frac{1}{i\sqrt{2}}\left( \frac{%
\partial }{\partial x}-x\right) ,  \label{mr7} \\
&&a=\frac{p_{x}-ix}{\sqrt{2}}=\frac{1}{i\sqrt{2}}\left( \frac{\partial }{%
\partial x}+x\right)  \label{mr8}
\end{eqnarray}%
with the familiar commutator $\left[ a,a^{\dagger }\right] $ $=aa^{\dagger
}-a^{\dagger }a=1$ \cite{Flu}. One can see that%
\begin{eqnarray}
&&a_{x}\psi =a_{x}F\left[ \chi \right] =F\left[ ia_{y}\chi \right] ,
\label{mr9} \\
&&a_{x}^{\dagger }\psi =a_{x}^{\dagger }F\left[ \chi \right] =F\left[
-ia_{y}^{\dagger }\chi \right] ,  \notag
\end{eqnarray}%
or%
\begin{equation}
a_{x}\rightarrow ia_{y},\qquad a_{x}^{\dagger }\rightarrow -ia_{y}^{\dagger }
\label{mr10}
\end{equation}%
under the Fourier transform. This observation will be important in the next
section.

Finally one can summarize all results on solution of the Cauchy initial
value problems for the modified oscillator under consideration in a form of
the commutative evolution diagram on Figure~3. We denote%
\begin{eqnarray}
U\left( t\right) \psi \left( x\right) &=&\int_{-\infty }^{\infty
}G_{U}\left( x,y,t\right) \ \psi \left( y\right) \ dy,  \label{mr11} \\
K\left( t\right) \psi \left( x\right) &=&\int_{-\infty }^{\infty
}K_{U}\left( x,y,t\right) \ \psi \left( y\right) \ dy,  \label{mr12} \\
V\left( t\right) \psi \left( x\right) &=&\int_{-\infty }^{\infty
}G_{V}\left( x,y,t\right) \ \psi \left( y\right) \ dy,  \label{mr13} \\
L\left( t\right) \psi \left( x\right) &=&\int_{-\infty }^{\infty
}K_{V}\left( x,y,t\right) \ \psi \left( y\right) \ dy.  \label{mr14}
\end{eqnarray}%
%
%
\begin{figure}[ptbh]
\centering\scalebox{0.75}{\includegraphics{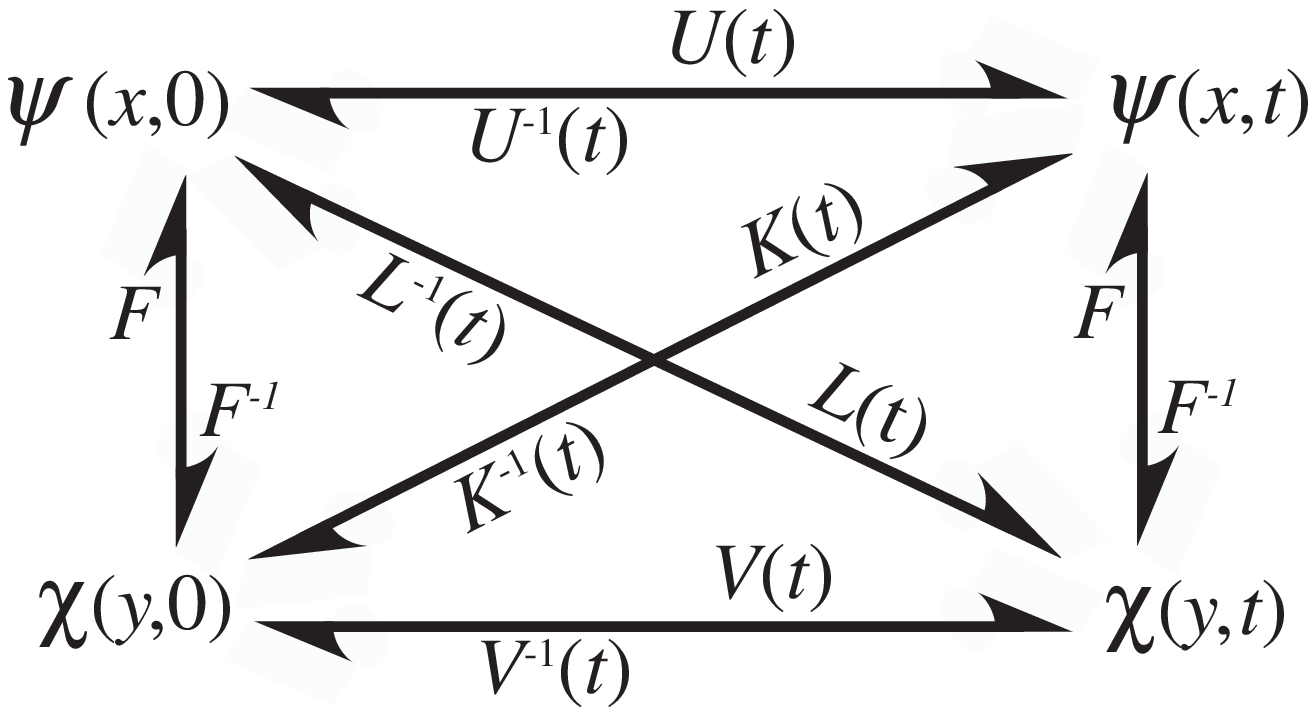}}
\caption{The commutative evolution diagram in $\boldsymbol{R}$.}
\end{figure}
%
\noindent The kernels of these integral operators are defined as follows.
Here $G_{U}\left( x,y,t\right) $ and $G_{V}\left( x,y,t\right) $ are the
Green functions in (\ref{in5}) and (\ref{sol23}), respectively. The kernels $%
K_{U}\left( x,y,t\right) $ and $K_{V}\left( x,y,t\right) $ are given by (\ref%
{sym21}) and (\ref{sym25}), respectively. The following operator identities
hold%
\begin{eqnarray}
&&U\left( t\right) =K\left( t\right) F^{-1}=FL\left( t\right) =FV\left(
t\right) F^{-1},  \label{mr15} \\
&&V\left( t\right) =L\left( t\right) F=F^{-1}K\left( t\right) =F^{-1}U\left(
t\right) F,  \label{mr16} \\
&&U^{-1}\left( t\right) =FK^{-1}\left( t\right) =L^{-1}\left( t\right)
F^{-1}=FV^{-1}\left( t\right) F^{-1},  \label{mr17} \\
&&V^{-1}\left( t\right) =F^{-1}L^{-1}\left( t\right) =K^{-1}\left( t\right)
F=F^{-1}U^{-1}\left( t\right) F,  \label{mr18} \\
&&K\left( t\right) =FL\left( t\right) F,\qquad L\left( t\right)
=F^{-1}K\left( t\right) F^{-1},  \label{mr19} \\
&&K^{-1}\left( t\right) =F^{-1}L^{-1}\left( t\right) F^{-1},\quad
L^{-1}\left( t\right) =FK^{-1}\left( t\right) F.  \label{mr20}
\end{eqnarray}%
Here $F$ and $F^{-1}$ are the operators of Fourier transform and its
inverse, respectively, which relate the wave functions in the coordinate and
momentum representations%
\begin{equation*}
\psi =F\left[ \chi \right] ,\qquad \chi =F^{-1}\left[ \psi \right]
\end{equation*}%
\noindent at any given moment of time thus representing the vertical arrows
at our diagram. The time evolution operators $U\left( t\right) ,$ $V\left(
t\right) $ and their inverses $U^{-1}\left( t\right) ,$ $V^{-1}\left(
t\right) $ correspond to the horizontal arrows in the coordinate and
momentum representations, respectively. They obey the symmetry with respect
to the time reversal, which has been discussed in section~5.

In order to discuss the diagonal arrows of the time evolution diagram on
Figure~3, we have to go back, say, to the particular solution (\ref{sym21}).
A more general solution of the Schr\"{o}dinger equation (\ref{sol1})--(\ref%
{sol2}) can be obtained by the superposition principle in the form%
\begin{equation}
\psi \left( x,t\right) =\int_{-\infty }^{\infty }K_{U}\left( x,y,t\right) \
\chi \left( y,0\right) \ dy,  \label{mr21}
\end{equation}%
where $\chi $ is an arbitrary function, independent of time, such that the
integral converges and one can interchange the differentiation and
integration. In view of the continuity of the kernel at $t=0,$ we get%
\begin{equation}
\psi \left( x,0\right) =\frac{1}{\sqrt{2\pi }}\int_{-\infty }^{\infty
}e^{ixy}\ \chi \left( y,0\right) \ dy,  \label{mr22}
\end{equation}%
which simply relates the initial data in the coordinate and momentum
representations. Then solution of the initial value problem is given by the
inverse of the Fourier transform%
\begin{equation}
\chi \left( y,0\right) =\frac{1}{\sqrt{2\pi }}\int_{-\infty }^{\infty
}e^{-ixy}\ \psi \left( x,0\right) \ dx  \label{mr23}
\end{equation}%
followed by the back substitution of this expression into (\ref{mr21}). This
implies the above factorization $U\left( t\right) =K\left( t\right) F^{-1}$
of the corresponding time evolution operator; see (\ref{mr15}). The Green
function (\ref{in5}) can be derived as%
\begin{equation}
G_{U}\left( x,y,t\right) =\frac{1}{\sqrt{2\pi }}\int_{-\infty }^{\infty
}K_{U}\left( x,z,t\right) \ e^{-iyz}\ dz  \label{mr24}
\end{equation}%
with the help of the integral (\ref{gauss}). The second equation, $U\left(
t\right) =FL\left( t\right) ,$ is related to following integral%
\begin{equation}
G_{U}\left( x,y,t\right) =\frac{1}{\sqrt{2\pi }}\int_{-\infty }^{\infty
}e^{ixz}\ K_{V}\left( z,y,t\right) \ dz.  \label{mr25}
\end{equation}%
The meaning of the operator $L\left( t\right) ,$ represented by another
diagonal arrow on the time evolution diagram, is established in a similar
fashion. One can see that the relation $V\left( t\right) =L\left( t\right) F$
in (\ref{mr16}) follows from the elementary integral%
\begin{equation}
G_{V}\left( x,y,t\right) =\frac{1}{\sqrt{2\pi }}\int_{-\infty }^{\infty
}K_{V}\left( x,z,t\right) \ e^{iyz}\ dz  \label{mr26}
\end{equation}%
and $K\left( t\right) =FV\left( t\right) $ corresponds to%
\begin{equation}
K_{U}\left( x,y,t\right) =\frac{1}{\sqrt{2\pi }}\int_{-\infty }^{\infty
}e^{ixz}\ G_{V}\left( z,y,t\right) \ dz.  \label{mr27}
\end{equation}%
This proves (\ref{mr15})--(\ref{mr16}). The inverses $K^{-1}\left( t\right) $
and $L^{-1}\left( t\right) $ are found, for instance, by the inverse of
Fourier transform similar to (\ref{fourier})--(\ref{fourierinv}). They are
not directly related to the reversal of time. We leave further details about
the structure of the commutative diagram on Figure~3 to the reader.

\section{The Case of $n$-Dimensions}

In the case of $\boldsymbol{R}^{n}$ with an arbitrary number of dimensions,
the Schr\"{o}dinger equation for a modified oscillator (\ref{in1}) with the
original Hamiltonian%
\begin{equation}
H\left( t\right) =\frac{1}{2}\sum_{s=1}^{n}\left( a_{s}a_{s}^{\dagger
}+a_{s}^{\dagger }a_{s}\right) +\frac{1}{2}e^{2it}\sum_{s=1}^{n}\left(
a_{s}\right) ^{2}+\frac{1}{2}e^{-2it}\sum_{s=1}^{n}\left( a_{s}^{\dagger
}\right) ^{2},  \label{nd1}
\end{equation}%
considered by Meiler, Cordero-Soto, and Suslov \cite{Me:Co:Su}, has the
Green function of the form
\begin{eqnarray}
G_{t}\left( \boldsymbol{x},\boldsymbol{x}^{\prime }\right)
&=&\dprod_{s=1}^{n}G_{t}\left( x_{s},x_{s}^{\prime }\right)  \label{nd2} \\
&=&\left( \frac{1}{2\pi i\left( \cos t\sinh t+\sin t\cosh t\right) }\right)
^{n/2}  \notag \\
&&\times \exp \left( \frac{\left( \boldsymbol{x}^{2}-\boldsymbol{x}^{\prime
2}\right) \sin t\sinh t+2\boldsymbol{x}\cdot \boldsymbol{x}^{\prime }-\left(
\boldsymbol{x}^{2}+\boldsymbol{x}^{\prime 2}\right) \cos t\cosh t}{2i\left(
\cos t\sinh t+\sin t\cosh t\right) }\right) .  \notag
\end{eqnarray}%
Solution of the Cauchy initial value problem can be written as%
\begin{equation}
\psi \left( \boldsymbol{x},t\right) =\int_{\boldsymbol{R}^{n}}G_{t}\left(
\boldsymbol{x},\boldsymbol{x}^{\prime }\right) \ \psi \left( \boldsymbol{x}%
^{\prime },0\right) \ d\boldsymbol{x}^{\prime },  \label{nd2a}
\end{equation}%
where $dv^{\prime }=d\boldsymbol{x}^{\prime }=dx_{1}^{\prime }\cdot ...\
\cdot dx_{n}^{\prime }.$ The propagator expansion in the hyperspherical
harmonics is given by%
\begin{equation}
G_{t}\left( \boldsymbol{x},\boldsymbol{x}^{\prime }\right) =\sum_{K\nu
}Y_{K\nu }\left( \Omega \right) \ Y_{K\nu }^{\ast }\left( \Omega ^{\prime
}\right) \ \mathcal{G}_{t}^{K}\left( r,r^{\prime }\right)  \label{nd3}
\end{equation}%
with%
\begin{eqnarray}
\mathcal{G}_{t}^{K}\left( r,r^{\prime }\right) &=&\frac{e^{-i\pi \left(
2K+n\right) /4}}{2^{K+n/2-1}\Gamma \left( K+n/2\right) }\ \frac{\left(
rr^{\prime }\right) ^{K}}{\left( \cos t\sinh t+\sin t\cosh t\right) ^{K+n/2}}
\label{nd4} \\
&&\times \exp \left( i\frac{\left( r^{2}+\left( r^{\prime }\right)
^{2}\right) \cos t\cosh t-\left( r^{2}-\left( r^{\prime }\right) ^{2}\right)
\sin t\sinh t}{2\left( \cos t\sinh t+\sin t\cosh t\right) }\right)  \notag \\
&&\times ~_{0}F_{1}\left(
\begin{array}{c}
- \\
K+n/2%
\end{array}%
;\ -\frac{\left( rr^{\prime }\right) ^{2}}{4\left( \cos t\sinh t+\sin t\cosh
t\right) ^{2}}\right) .  \notag
\end{eqnarray}%
Here $Y_{K\nu }\left( \Omega \right) $ are the hyperspherical harmonics
constructed by the given tree $T$ in the graphical approach of Vilenkin,
Kuznetsov and Smorodinski\u{\i} \cite{Ni:Su:Uv}, the integer $K$ corresponds
to the constant of separation of the variables at the root of $T$ (denoted
by $K$ due to the tradition of the method of $K$-harmonics in nuclear
physics \cite{Smir:Shit}) and $\nu =\left\{ l_{1},l_{2},...\ ,l_{p}\right\} $
is the set of all other subscripts corresponding to the remaining vertexes
of the binary tree $T.$ These formulas imply the familiar expansion of a
plane wave in $\boldsymbol{R}^{n}$ in terms of the hyperspherical harmonics%
\begin{equation}
e^{i\boldsymbol{x}\cdot \boldsymbol{x}^{\prime }}=rr^{\prime }\left( \frac{%
2\pi }{rr^{\prime }}\right) ^{n/2}\sum_{K\nu }i^{K}\ Y_{K\nu }^{\ast }\left(
\Omega \right) \ Y_{K\nu }\left( \Omega ^{\prime }\right) \
J_{K+n/2-1}\left( rr^{\prime }\right) ,  \label{exp1}
\end{equation}%
where%
\begin{equation}
J_{\mu }\left( z\right) =\frac{\left( z/2\right) ^{\mu }}{\Gamma \left( \mu
+1\right) }\ ~_{0}F_{1}\left(
\begin{array}{c}
- \\
\mu +1%
\end{array}%
;\ -\frac{z^{2}}{4}\right)  \label{exp2}
\end{equation}%
is the Bessel function. See \cite{Me:Co:Su} and references therein for more
details. It is worth noting that the Green function (\ref{in5}) was
originally found by Meiler, Cordero-Soto, and Suslov as the special case $%
n=1 $ of the expansion (\ref{nd3})--(\ref{nd4}). The dynamical $SU\left(
1,1\right) $ symmetry of the harmonic oscillator wave functions, Bargmann's
functions for the discrete positive series of the irreducible
representations of this group, the Fourier integral of a weighted product of
the Meixner--Pollaczek polynomials, a Hankel-type integral transform and the
hyperspherical harmonics were utilized in order to derive the $n$%
-dimensional Green function.

Our results show that the \textquotedblleft dual\textquotedblright\ Schr\"{o}%
dinger equation (\ref{in6}) with a new Hamiltonian of the form%
\begin{equation}
H\left( \tau \right) =\frac{1}{2}\sum_{s=1}^{n}\left( a_{s}a_{s}^{\dagger
}+a_{s}^{\dagger }a_{s}\right) +\frac{1}{2}e^{-i\arctan \left( 2\tau \right)
}\sum_{s=1}^{n}\left( a_{s}\right) ^{2}+\frac{1}{2}e^{i\arctan \left( 2\tau
\right) }\sum_{s=1}^{n}\left( a_{s}^{\dagger }\right) ^{2}  \label{nd5}
\end{equation}%
has the propagator that is almost identical to (\ref{nd2}) but with $%
\boldsymbol{x}\leftrightarrow \boldsymbol{x}^{\prime }.$ Indeed, in the case
of $n$-dimensions one has%
\begin{equation}
H\left( \tau \right) =\sum_{s=1}^{n}H_{s}\left( \tau \right) ,  \label{nd6}
\end{equation}%
where we denote%
\begin{equation}
H_{s}\left( \tau \right) =\frac{1}{2}\left( a_{s}a_{s}^{\dagger
}+a_{s}^{\dagger }a_{s}\right) +\frac{1}{2}e^{-i\arctan \left( 2\tau \right)
}\left( a_{s}\right) ^{2}+\frac{1}{2}e^{i\arctan \left( 2\tau \right)
}\left( a_{s}^{\dagger }\right) ^{2}.  \label{nd7}
\end{equation}%
If one chooses%
\begin{eqnarray}
\psi _{s} &=&\psi _{s}\left( x_{s},t\right) =G_{t}\left( x_{s}^{\prime
},x_{s}\right)  \label{nd8} \\
&=&\left( \frac{1}{2\pi i\left( \cos t\sinh t+\sin t\cosh t\right) }\right)
^{1/2}  \notag \\
&&\times \exp \left( \frac{\left( x_{s}^{\prime 2}-x_{s}^{2}\right) \sin
t\sinh t+2x_{s}^{\prime }x_{s}-\left( x_{s}^{\prime 2}+x_{s}^{2}\right) \cos
t\cosh t}{2i\left( \cos t\sinh t+\sin t\cosh t\right) }\right)  \notag
\end{eqnarray}%
with%
\begin{equation}
i\frac{\partial \psi _{s}}{\partial \tau }=H_{s}\left( \tau \right) \psi
_{s},\qquad t=\frac{1}{2}\sinh \left( 2\tau \right)  \label{nd8a}
\end{equation}%
and denotes%
\begin{equation}
\psi =\dprod_{k=1}^{n}\psi _{k}=\dprod_{k=1}^{n}G_{t}\left( x_{k}^{\prime
},x_{k}\right) =G_{t}\left( \boldsymbol{x}^{\prime },\boldsymbol{x}\right) ,
\label{nd9}
\end{equation}%
then%
\begin{equation}
i\frac{\partial \psi }{\partial \tau }=\sum_{s=1}^{n}\left( i\frac{\partial
\psi _{s}}{\partial \tau }\right) \dprod_{k\neq s}\psi _{k}  \label{nd10}
\end{equation}%
and%
\begin{equation}
H\left( \tau \right) \psi =\sum_{s=1}^{n}\left( H_{s}\left( \tau \right)
\psi _{s}\right) \dprod_{k\neq s}\psi _{k}.  \label{nd11}
\end{equation}%
As a result,%
\begin{equation}
\left( i\frac{\partial }{\partial \tau }-H\left( \tau \right) \right) \psi
=\sum_{s=1}^{n}\left( i\frac{\partial \psi _{s}}{\partial \tau }-H_{s}\left(
\tau \right) \psi _{s}\right) \dprod_{k\neq s}\psi _{k}\equiv 0,
\label{nd12}
\end{equation}%
and Eq.~(\ref{in6}) for the $n$-dimensional propagator is satisfied. For the
initial data, formally,%
\begin{equation}
\lim_{t\rightarrow 0^{+}}G_{t}\left( \boldsymbol{x}^{\prime },\boldsymbol{x}%
\right) =\dprod_{k=1}^{n}\lim_{t\rightarrow 0^{+}}G_{t}\left( x_{k}^{\prime
},x_{k}\right) =\dprod_{k=1}^{n}\delta \left( x_{k}^{\prime }-x_{k}\right)
=\delta \left( \boldsymbol{x}-\boldsymbol{x}^{\prime }\right) ,  \label{nd13}
\end{equation}%
where $\delta \left( \boldsymbol{x}\right) $ is the Dirac delta function in $%
\boldsymbol{R}^{n}.$ Further details are left to the reader.

The $n$-dimensional version of the Hamiltonian corresponding to the
coefficients (\ref{sol18}) is given by%
\begin{equation}
H\left( t\right) =\frac{1}{2}\sum_{s=1}^{n}\left( a_{s}a_{s}^{\dagger
}+a_{s}^{\dagger }a_{s}\right) -\frac{1}{2}e^{2it}\sum_{s=1}^{n}\left(
a_{s}\right) ^{2}-\frac{1}{2}e^{-2it}\sum_{s=1}^{n}\left( a_{s}^{\dagger
}\right) ^{2}  \label{nd14}
\end{equation}%
with the propagator%
\begin{eqnarray}
G_{t}\left( \boldsymbol{x},\boldsymbol{x}^{\prime }\right) &=&\left( \frac{1%
}{2\pi i\left( \sin t\cosh t-\cos t\sinh t\right) }\right) ^{n/2}
\label{nd15} \\
&&\times \exp \left( \frac{\left( \boldsymbol{x}^{2}+\boldsymbol{x}^{\prime
2}\right) \cos t\cosh t-2\boldsymbol{x}\cdot \boldsymbol{x}^{\prime }+\left(
\boldsymbol{x}^{2}-\boldsymbol{x}^{\prime 2}\right) \sin t\sinh t}{2i\left(
\cos t\sinh t-\sin t\cosh t\right) }\right) .  \notag
\end{eqnarray}%
The dual counterpart of this Hamiltonian with respect to time reversal has
the form%
\begin{equation}
H\left( \tau \right) =\frac{1}{2}\sum_{s=1}^{n}\left( a_{s}a_{s}^{\dagger
}+a_{s}^{\dagger }a_{s}\right) -\frac{1}{2}e^{-i\arctan \left( 2\tau \right)
}\sum_{s=1}^{n}\left( a_{s}\right) ^{2}-\frac{1}{2}e^{i\arctan \left( 2\tau
\right) }\sum_{s=1}^{n}\left( a_{s}^{\dagger }\right) ^{2}  \label{nd16}
\end{equation}%
and one has to interchange $\boldsymbol{x}\leftrightarrow \boldsymbol{x}%
^{\prime }$ in (\ref{nd15}) in order to obtain the corresponding Green
function. It is worth noting that the Hamiltonians (\ref{nd1}) and (\ref%
{nd14}) (respectively, (\ref{nd5}) and (\ref{nd16})) are transforming into
each other under the substitution $a_{s}\rightarrow ia_{s},$ $a_{s}^{\dagger
}\rightarrow -ia_{s}^{\dagger },$ which preserves the commutation relations
of the creation and annihilation operators. As we have seen in the previous
section this property is related to solving the problem in the coordinate
and momentum representations.

Combining all four cases together, one may summarize that two Hamiltonians,%
\begin{equation}
H_{\pm }\left( t\right) =\frac{1}{2}\sum_{s=1}^{n}\left( a_{s}a_{s}^{\dagger
}+a_{s}^{\dagger }a_{s}\right) \pm \frac{1}{2}e^{2it}\sum_{s=1}^{n}\left(
a_{s}\right) ^{2}\pm \frac{1}{2}e^{-2it}\sum_{s=1}^{n}\left( a_{s}^{\dagger
}\right) ^{2},  \label{nd17}
\end{equation}%
and their duals with respect to the time reversal,%
\begin{equation}
H_{\pm }\left( \tau \right) =\frac{1}{2}\sum_{s=1}^{n}\left(
a_{s}a_{s}^{\dagger }+a_{s}^{\dagger }a_{s}\right) \pm \frac{1}{2}%
e^{-i\arctan \left( 2\tau \right) }\sum_{s=1}^{n}\left( a_{s}\right) ^{2}\pm
\frac{1}{2}e^{i\arctan \left( 2\tau \right) }\sum_{s=1}^{n}\left(
a_{s}^{\dagger }\right) ^{2}  \label{nd18}
\end{equation}%
with $\tau =\frac{1}{2}\sinh \left( 2t\right) ,$ have the following Green
functions:%
\begin{eqnarray}
G_{t}^{\pm }\left( \boldsymbol{x},\boldsymbol{x}^{\prime }\right) &=&\left(
\frac{1}{2\pi i\left( \sin t\cosh t\pm \cos t\sinh t\right) }\right) ^{n/2}
\label{nd19} \\
&&\times \exp \left( \frac{\pm \left( \boldsymbol{x}^{2}-\boldsymbol{x}%
^{\prime 2}\right) \sin t\sinh t+2\boldsymbol{x}\cdot \boldsymbol{x}^{\prime
}-\left( \boldsymbol{x}^{2}+\boldsymbol{x}^{\prime 2}\right) \cos t\cosh t}{%
2i\left( \sin t\cosh t\pm \cos t\sinh t\right) }\right) .  \notag
\end{eqnarray}%
This expression is valid for two Hamiltonians (\ref{nd17}), respectively.
One has to interchange $\boldsymbol{x}\leftrightarrow \boldsymbol{x}^{\prime
}$ for the case of the dual Hamiltonians (\ref{nd18}).

In a similar fashion, the $n$-dimensional form of the kernels (\ref{sym21})
and (\ref{sym25}) is%
\begin{eqnarray}
K_{t}^{\pm }\left( \boldsymbol{x},\boldsymbol{x}^{\prime }\right) &=&\left(
\frac{1}{2\pi \left( \cos t\cosh t\pm \sin t\sinh t\right) }\right) ^{n/2}
\label{nd20} \\
&&\times \exp \left( \frac{\left( \boldsymbol{x}^{2}+\boldsymbol{x}^{\prime
2}\right) \sin t\cosh t\mp 2\boldsymbol{x}\cdot \boldsymbol{x}^{\prime }\mp
\left( \boldsymbol{x}^{2}-\boldsymbol{x}^{\prime 2}\right) \cos t\sinh t}{%
2i\left( \cos t\cosh t\pm \sin t\sinh t\right) }\right)  \notag
\end{eqnarray}%
and%
\begin{eqnarray}
G_{t}^{\pm }\left( \boldsymbol{x},\boldsymbol{x}^{\prime }\right) &=&\frac{1%
}{\left( 2\pi \right) ^{n/2}}\int_{\boldsymbol{R}^{n}}K_{t}^{\pm }\left(
\boldsymbol{x},\boldsymbol{x}^{\prime \prime }\right) \ e^{\mp i\boldsymbol{x%
}^{\prime }\cdot \boldsymbol{x}^{\prime \prime }}\ d\boldsymbol{x}^{\prime
\prime }  \label{nd21} \\
&=&\frac{1}{\left( 2\pi \right) ^{n/2}}\int_{\boldsymbol{R}^{n}}e^{\pm i%
\boldsymbol{x}\cdot \boldsymbol{x}^{\prime \prime }}\ K_{t}^{\mp }\left(
\boldsymbol{x}^{\prime \prime },\boldsymbol{x}^{\prime }\right) \ d%
\boldsymbol{x}^{\prime \prime }.  \notag
\end{eqnarray}%
%
%
\begin{figure}[ptbh]
\centering\scalebox{.75}{\includegraphics{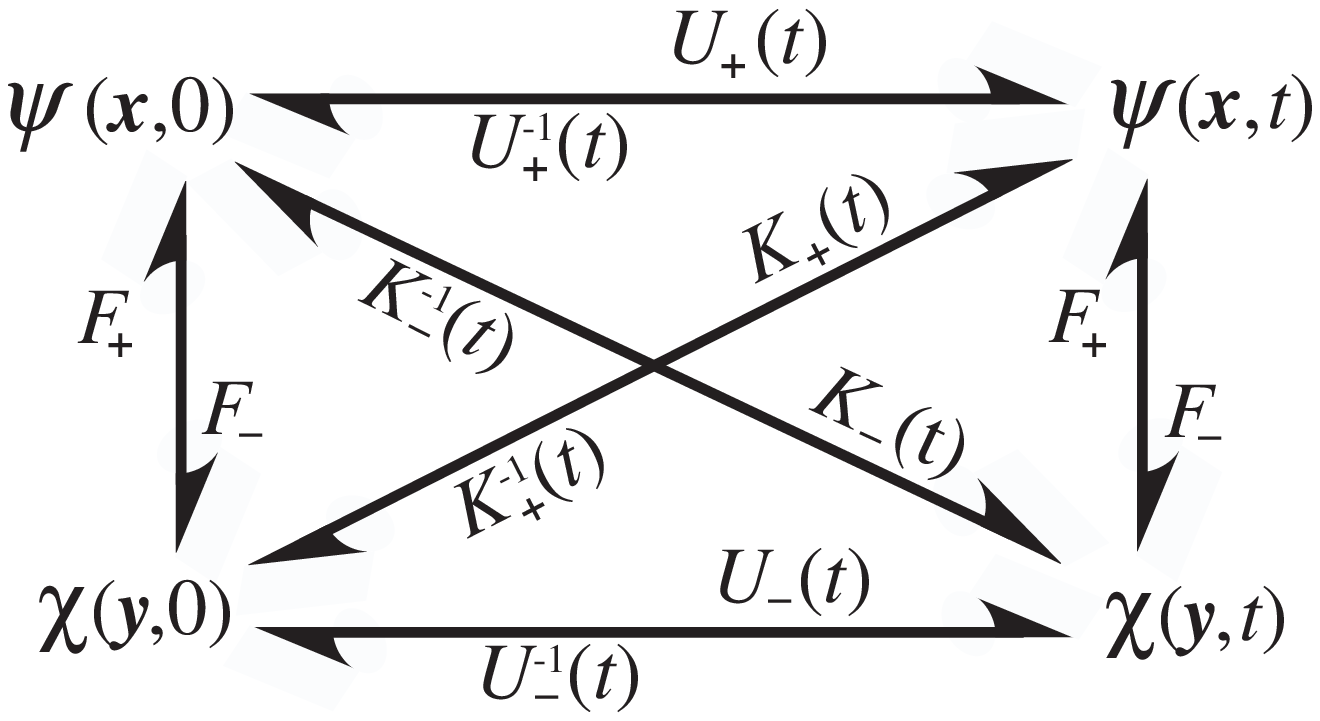}}
\caption{The commutative evolution diagram in ${\boldsymbol{R}}^{n}.$}
\end{figure}
%
\noindent Denoting%
\begin{eqnarray}
U_{\pm }\left( t\right) \psi \left( \boldsymbol{x}\right) &=&\int_{%
\boldsymbol{R}^{n}}G_{t}^{\pm }\left( \boldsymbol{x},\boldsymbol{x}^{\prime
}\right) \ \psi \left( \boldsymbol{x}^{\prime }\right) \ d\boldsymbol{x}%
^{\prime },  \label{nd22} \\
K_{\pm }\left( t\right) \psi \left( \boldsymbol{x}\right) &=&\int_{%
\boldsymbol{R}^{n}}K_{t}^{\pm }\left( \boldsymbol{x},\boldsymbol{x}^{\prime
}\right) \ \psi \left( \boldsymbol{x}^{\prime }\right) \ d\boldsymbol{x}%
^{\prime },  \notag
\end{eqnarray}%
and%
\begin{equation}
F_{\pm }\psi \left( \boldsymbol{x}\right) =\frac{1}{\left( 2\pi \right)
^{n/2}}\int_{\boldsymbol{R}^{n}}e^{\pm i\boldsymbol{x}\cdot \boldsymbol{x}%
^{\prime }}\ \psi \left( \boldsymbol{x}^{\prime }\right) \ d\boldsymbol{x}%
^{\prime }  \label{nd23}
\end{equation}%
one arrives at the commutative evolution diagram in $\boldsymbol{R}^{n}$ on
Figure~4. The corresponding relations are%
\begin{eqnarray}
&&U_{\pm }\left( t\right) =K_{\pm }\left( t\right) F_{\mp }=F_{\pm }K_{\mp
}\left( t\right) =F_{\pm }U_{\mp }\left( t\right) F_{\mp },  \label{nd24} \\
&&U_{\pm }^{-1}\left( t\right) =F_{\pm }K_{\pm }^{-1}\left( t\right) =K_{\mp
}^{-1}\left( t\right) F_{\mp }=F_{\pm }U_{\mp }^{-1}\left( t\right) F_{\mp }.
\notag
\end{eqnarray}%
We leave further details to the reader.

A certain time-dependent Schr\"{o}dinger equation with variable coefficients
was considered in \cite{Me:Co:Su} in a pure algebraic manner in connection
with representations of the group $SU\left( 1,1\right) $ in an abstract
Hilbert space. Our Hamiltonians (\ref{nd17}) and (\ref{nd18}) belong to the
same class thus providing new explicit realizations of this model in
addition to several cases already discussed by Meiler, Cordero-Soto, and
Suslov.

\section{Eigenfunction Expansions}

The normalized wave functions of the $n$-dimensional harmonic oscillator%
\begin{equation}
H_{0}\Psi =E\Psi ,\qquad H_{0}=\frac{1}{2}\sum_{s=1}^{n}\left( -\frac{%
\partial ^{2}}{\partial x_{s}^{2}}+x_{s}^{2}\right)  \label{h0}
\end{equation}%
have the form%
\begin{equation}
\Psi \left( \boldsymbol{x}\right) =\Psi _{NK\nu }\left( r,\Omega \right)
=Y_{K\nu }\left( \Omega \right) \ R_{NK}\left( r\right) ,  \label{h1}
\end{equation}%
where $Y_{K\nu }\left( \Omega \right) $ are the hyperspherical harmonics
associated with a binary tree $T,$ the integer number $K$ corresponds to the
constant of separation of the variables at the root of $T$ and $\nu =\left\{
l_{1},l_{2},...\ ,l_{p}\right\} $ is the set of all other subscripts
corresponding to the remaining vertexes of the binary tree $T;$ see \cite%
{Ni:Su:Uv}, \cite{Smir:Shit}, \cite{Vil} for a graphical approach of
Vilenkin, Kuznetsov and Smorodinski\u{\i} to the theory of spherical
harmonics. The radial functions are given by%
\begin{equation}
R_{NK}\left( r\right) =\sqrt{\frac{2\left[ \left( N-K\right) /2\right] !}{%
\Gamma \left[ \left( N+K+n\right) /2\right] }}\ \exp \left( -r^{2}/2\right)
\ r^{K}\ L_{\left( N-K\right) /2}^{K+n/2-1}\left( r^{2}\right) ,  \label{h2}
\end{equation}%
where $L_{k}^{\alpha }\left( \xi \right) $ are the Laguerre polynomials. The
corresponding energy levels are%
\begin{equation}
E=E_{N}=N+n/2,\qquad \left( N-K\right) /2=k=0,1,2,...\   \label{h3}
\end{equation}%
and we can use the $SU\left( 1,1\right) $-notation for the wave function as
follows%
\begin{equation}
\psi _{jm\left\{ \nu \right\} }\left( \boldsymbol{x}\right) =\Psi _{NK\nu
}\left( r,\Omega \right) =Y_{K\nu }\left( \Omega \right) \ R_{NK}\left(
r\right) ,  \label{h4}
\end{equation}%
where the new quantum numbers are given by $j=K/2+n/4-1$ and $m=N/2+n/4$
with $m=j+1,$ $j+2,...\ .$ The inequality $m\geq j+1$ holds because of the
quantization rule (\ref{h3}), which gives $N=K,$ $K+2,$ $K+4,...\ $ $.$ See
\cite{Me:Co:Su}, \cite{Ni:Su:Uv} and \cite{Smir:Shit} for more details on
the group theoretical properties of the $n$-dimensional harmonic oscillator
wave functions.

The Cauchy initial value problem for the Schr\"{o}dinger equation (\ref{in1}%
) with the Hamiltonian of a modified oscillator (\ref{nd1}) has also the
eigenfunction expansion form of the solution \cite{Me:Co:Su}:

\begin{equation}
\psi \left( \boldsymbol{x},t\right) =\sum_{j\left\{ \nu \right\}
}\sum_{m=j+1}^{\infty }c_{m}\left( t\right) \ \psi _{jm\left\{ \nu \right\}
}\left( \boldsymbol{x}\right)  \label{h5}
\end{equation}%
with the time dependent coefficients%
\begin{equation}
c_{m}\left( t\right) =e^{-2imt}\sum_{m^{\prime }=j+1}^{\infty }i^{m^{\prime
}-m}v_{m^{\prime }m}^{j}\left( 2t\right) \ \int_{\boldsymbol{R}^{n}}\psi
_{jm^{\prime }\left\{ \nu \right\} }^{\ast }\left( \boldsymbol{x}^{\prime
}\right) \psi \left( \boldsymbol{x}^{\prime },0\right) \ dv^{\prime }
\label{h6}
\end{equation}%
given in terms of the Bargmann functions \cite{Bargmann47}, \cite{Ni:Su:Uv}
and \cite{Vil}%
\begin{align}
& v_{mm^{\prime }}^{j}\left( \mu \right) =\frac{\left( -1\right) ^{m-j-1}}{%
\Gamma \left( 2j+2\right) }\sqrt{\frac{\left( m+j\right) !\left( m^{\prime
}+j\right) !}{\left( m-j-1\right) !\left( m^{\prime }-j-1\right) !}}\ \left(
\sinh \frac{\mu }{2}\right) ^{-2j-2}\left( \tanh \frac{\mu }{2}\right)
^{m+m^{\prime }}  \notag \\
& \ \ \ \ \qquad \qquad \times \ \ _{2}F_{1}\left(
\begin{array}{c}
-m+j+1,\ -m^{\prime }+j+1 \\
2j+2%
\end{array}%
;\ -\frac{1}{\sinh ^{2}\left( \mu /2\right) }\right) .  \label{h7}
\end{align}%
Choosing the initial data in (\ref{nd2a}) and (\ref{h5})--(\ref{h6}) as $%
\psi \left( \boldsymbol{x},0\right) =\delta \left( \boldsymbol{x}-%
\boldsymbol{x}^{\prime }\right) ,$ we arrive at the eigenfunction expansion
for the Green function%
\begin{equation}
G_{t}\left( \boldsymbol{x},\boldsymbol{x}^{\prime }\right) =\sum_{j\left\{
\nu \right\} }\sum_{m,m^{\prime }=j+1}^{\infty }e^{-2imt}\ i^{m^{\prime
}-m}\ v_{m^{\prime }m}^{j}\left( 2t\right) \ \psi _{jm\left\{ \nu \right\}
}\left( \boldsymbol{x}\right) \psi _{jm^{\prime }\left\{ \nu \right\}
}^{\ast }\left( \boldsymbol{x}^{\prime }\right) ,  \label{h8}
\end{equation}%
where by (\ref{nd15}) the following symmetry property holds%
\begin{equation}
G_{t}\left( \boldsymbol{x},\boldsymbol{x}^{\prime }\right) =G_{-t}^{\ast
}\left( \boldsymbol{x},\boldsymbol{x}^{\prime }\right) .  \label{h9}
\end{equation}

In this paper we have found solution of the Cauchy initial value problem for
the new Hamiltonian (\ref{nd5}) in an integral form%
\begin{equation}
\psi \left( \boldsymbol{x},t\right) =\int_{\boldsymbol{R}^{n}}G_{t}\left(
\boldsymbol{x}^{\prime },\boldsymbol{x}\right) \ \psi \left( \boldsymbol{x}%
^{\prime },0\right) \ d\boldsymbol{x}^{\prime }.  \label{h10}
\end{equation}%
In view of (\ref{h8})--(\ref{h9}), the eigenfunction expansion of this
solution is given by%
\begin{equation}
\psi \left( \boldsymbol{x},t\right) =\sum_{j\left\{ \nu \right\}
}\sum_{m=j+1}^{\infty }c_{m}\left( t\right) \ \psi _{jm\left\{ \nu \right\}
}\left( \boldsymbol{x}\right) ,  \label{h11}
\end{equation}%
where%
\begin{equation}
c_{m}\left( t\right) =\sum_{m^{\prime }=j+1}^{\infty }\left( -i\right)
^{m-m^{\prime }}e^{-2im^{\prime }t}\left( v_{m^{\prime }m}^{j}\left(
-2t\right) \right) ^{\ast }\ \int_{\boldsymbol{R}^{n}}\psi _{jm^{\prime
}\left\{ \nu \right\} }^{\ast }\left( \boldsymbol{x}^{\prime }\right) \psi
\left( \boldsymbol{x}^{\prime },0\right) \ dv^{\prime }.  \label{h12}
\end{equation}%
This expansion is in agreement with the unitary infinite matrix of the
inverse operator in the basis of the harmonic oscillator wave functions; see
section~5.

The cases of the Hamiltonian (\ref{nd14}) and its dual (\ref{nd16}) can be
investigated by taking the Fourier transform of the expansions (\ref{h5})--(%
\ref{h6}) and (\ref{h11})--(\ref{h12}), respectively. The corresponding
transformations of the oscillator wave functions are%
\begin{equation}
i^{\pm N}\Psi _{NK\nu }\left( \boldsymbol{x}\right) =\frac{1}{\left( 2\pi
\right) ^{n/2}}\int_{\boldsymbol{R}^{n}}e^{\pm i\boldsymbol{x}\cdot
\boldsymbol{x}^{\prime }}\ \Psi _{NK\nu }\left( \boldsymbol{x}^{\prime
}\right) \ d\boldsymbol{x}^{\prime }.  \label{h13}
\end{equation}%
This can be evaluated in hyperspherical coordinates with the help of
expansion (\ref{exp1})--(\ref{exp2}), or by adding the $SU\left( 1,1\right) $%
-momenta according to the tree $T$ \cite{Ni:Su:Uv} and using linearity of
the Fourier transform. One can use%
\begin{equation}
e^{i\boldsymbol{x}\cdot \boldsymbol{x}^{\prime }}=\left( 2\pi \right)
^{n/2}\sum_{K\nu }i^{K}\ Y_{K\nu }^{\ast }\left( \Omega \right) \ Y_{K\nu
}\left( \Omega ^{\prime }\right) \ S_{-1}\left( r,r^{\prime }\right) \label%
{h14}
\end{equation}%
with%
\begin{equation}
S_{-1}\left( r,r^{\prime }\right) =\frac{\left( rr^{\prime }\right) ^{K}}{%
2^{K+n/2-1}\Gamma \left( K+n/2\right) }\ ~_{0}F_{1}\left(
\begin{array}{c}
- \\
K+n/2%
\end{array}%
;\ -\frac{\left( rr^{\prime }\right) ^{2}}{4}\right) \label{h15}
\end{equation}%
and%
\begin{equation}
\left( -1\right) ^{\left( N-K\right) /2}R_{NK}\left( r\right)
=\int_{0}^{\infty }S_{-1}\left( r,r^{\prime }\right) \ R_{NK}\left(
r^{\prime }\right) \left( r^{\prime }\right) ^{n-1}dr^{\prime }\label{h16}
\end{equation}%
as a special case of Eqs.~(7.3) and (7.6) of Ref.~\cite{Me:Co:Su} together with
the orthogonality property of hyperspherical harmonics. We leave
further details to the reader.

\section{Particular Solutions of The Nonlinear Schr\"{o}dinger Equations}

The method of solving the equation (\ref{sol1}) is extended in \cite%
{Cor-Sot:Lop:Sua:Sus} to the nonlinear Schr\"{o}dinger equation of\ the form%
\begin{equation}
i\frac{\partial \psi }{\partial t}=-a\left( t\right) \frac{\partial ^{2}\psi
}{\partial x^{2}}+b\left( t\right) x^{2}\psi -i\left( c\left( t\right) x%
\frac{\partial \psi }{\partial x}+d\left( t\right) \psi \right) +h\left(
t\right) \left\vert \psi \right\vert ^{2s}\psi ,\qquad s\geq 0.  \label{ns1}
\end{equation}%
We elaborate first on two cases (\ref{sol2}) and (\ref{sol3}). A particular
solution takes the form%
\begin{equation}
\psi =\psi \left( x,t\right) =K_{h}\left( x,y,t\right) =\frac{e^{i\phi }}{%
\sqrt{\mu \left( t\right) }}\ e^{i\left( \alpha \left( t\right) x^{2}+\beta
\left( t\right) xy+\gamma \left( t\right) y^{2}+\kappa \left( t\right)
\right) },\qquad \phi =\text{constant},  \label{ns2}
\end{equation}%
where equations (\ref{sol5})--(\ref{sol7}) hold and, in addition,%
\begin{equation}
\frac{d\kappa }{dt}=-\frac{h\left( t\right) }{\mu ^{s}\left( t\right) }%
,\qquad \kappa \left( t\right) =\kappa \left( 0\right) -\int_{0}^{t}\frac{%
h\left( \tau \right) }{\mu ^{s}\left( \tau \right) }\ d\tau ,  \label{ns3}
\end{equation}%
provided that the integral converges.

In the first case (\ref{sol2}), by the superposition principle, the general
solution of the characteristic equation (\ref{sol11}) has the form%
\begin{eqnarray}
\mu &=&c_{1}\mu _{1}\left( t\right) +c_{2}\mu _{2}\left( t\right)
\label{ns4} \\
&=&\cos t\left( c_{1}\cosh t+c_{2}\sinh t\right) +\sin t\left( c_{1}\sinh
t+c_{2}\cosh t\right)  \notag \\
&=&\frac{c_{1}+c_{2}}{\sqrt{2}}\ e^{t}\sin \left( t+\frac{\pi }{4}\right) +%
\frac{c_{1}-c_{2}}{\sqrt{2}}\ e^{-t}\cos \left( t+\frac{\pi }{4}\right)
\notag
\end{eqnarray}%
with $\mu ^{\prime }=2\cos t\left( c_{1}\sinh t+c_{2}\cosh t\right) $ and
\begin{equation}
\mu \left( 0\right) =c_{1},\qquad \mu ^{\prime }\left( 0\right) =2c_{2}.
\label{ns5}
\end{equation}%
Then%
\begin{eqnarray}
\alpha \left( t\right) &=&\frac{\cos t\left( c_{1}\sinh t+c_{2}\cosh
t\right) -\sin t\left( c_{1}\cosh t+c_{2}\sinh t\right) }{2\left( \cos
t\left( c_{1}\cosh t+c_{2}\sinh t\right) +\sin t\left( c_{1}\sinh
t+c_{2}\cosh t\right) \right) },  \label{ns6} \\
\beta \left( t\right) &=&\frac{c_{1}\beta \left( 0\right) }{\cos t\left(
c_{1}\cosh t+c_{2}\sinh t\right) +\sin t\left( c_{1}\sinh t+c_{2}\cosh
t\right) },  \label{ns7}
\end{eqnarray}%
and%
\begin{equation}
\gamma \left( t\right) =\gamma \left( 0\right) -\frac{c_{1}\beta ^{2}\left(
0\right) \left( \cos t\sinh t+\sin t\cosh t\right) }{2\left( \cos t\left(
c_{1}\cosh t+c_{2}\sinh t\right) +\sin t\left( c_{1}\sinh t+c_{2}\cosh
t\right) \right) }  \label{ns8}
\end{equation}%
as a result of elementary but somewhat tedious calculations. The first two
equations follow directly from (\ref{sol5}) and a constant multiple of the
first equation~(\ref{sol6}), respectively. One should use%
\begin{equation}
\frac{d\gamma }{dt}+a\left( t\right) \beta ^{2}=0,  \label{ns8a}
\end{equation}%
see \cite{Cor-Sot:Lop:Sua:Sus}, integration by parts as in (\ref{sol7}), and
an elementary integral
\begin{equation}
\int \frac{dt}{\left( c_{1}\sinh t+c_{2}\cosh t\right) ^{2}}=\frac{\sinh t}{%
c_{2}\left( c_{1}\sinh t+c_{2}\cosh t\right) }+C  \label{ns8b}
\end{equation}%
in order to derive (\ref{ns8}).

Two special cases are as follows. The original propagator (\ref{in5})
appears in the limit $c_{1}\rightarrow 0$ when $\beta \left( 0\right)
=-\left( c_{1}\right) ^{-1}$ and $\gamma \left( 0\right) =\left(
2c_{1}c_{2}\right) ^{-1}.$ The solution with the standing wave initial data $%
\psi \left( x,0\right) =e^{ixy}$ found in \cite{Cor-Sot:Lop:Sua:Sus}
corresponds to $c_{1}=1$ and $c_{2}=0.$

Equation (\ref{ns3}) can be explicitly integrated in some special cases,
say, when $h\left( t\right) =\lambda \mu ^{\prime }\left( t\right) :$%
\begin{equation}
\kappa \left( t\right) =\left\{
\begin{array}{ll}
\kappa \left( 0\right) -\dfrac{\lambda }{1-s}\left( \mu ^{1-s}\left(
t\right) -\mu ^{1-s}\left( 0\right) \right) , & \text{when }s\neq 1,\bigskip
\\
\kappa \left( 0\right) -\lambda \ln \left( \dfrac{\mu \left( t\right) }{\mu
\left( 0\right) }\right) , & \text{when }s=1.%
\end{array}%
\right.  \label{ns9}
\end{equation}%
Here $\mu \left( 0\right) \neq 0;$ cf.~\cite{Cor-Sot:Lop:Sua:Sus}. One may
treat the general particular solution of the form (\ref{ns2}) with the
coefficients (\ref{ns6})--(\ref{ns8}) and (\ref{ns9}) as an example of
application of yet unknown \textquotedblleft nonlinear\textquotedblright\
superposition principle for the Schr\"{o}dinger equation under consideration
for two particular solutions of a similar form with $c_{1}\neq 0,$ $c_{2}=0$
and $c_{1}=0,$ $c_{2}\neq 0.$

It is worth noting that function (\ref{ns2}) with the coefficients given by (%
\ref{ns4})--(\ref{ns9}) does also satisfy the following linear Schr\"{o}%
dinger equation%
\begin{equation}
i\frac{\partial \psi }{\partial t}=-a\left( t\right) \frac{\partial ^{2}\psi
}{\partial x^{2}}+b\left( t\right) x^{2}\psi -i\left( c\left( t\right) x%
\frac{\partial \psi }{\partial x}+d\left( t\right) \psi \right) +\frac{%
h\left( t\right) }{\mu ^{s}\left( t\right) }\ \psi ,\qquad s\geq 0.
\label{ns10}
\end{equation}%
Then a more general solution of this equation can be obtained by the
superposition principle as follows%
\begin{equation}
\psi \left( x,t\right) =\int_{-\infty }^{\infty }K_{h}\left( x,y,t\right) \
\chi \left( y\right) \ dy,  \label{ns11}
\end{equation}%
where $\chi $ is an arbitrary function such that the integral converges and
one can interchange differentiation and integration. Solution of the Cauchy
initial value problem simply requires an inversion of the integral%
\begin{equation}
\psi \left( x,0\right) =\int_{-\infty }^{\infty }K_{h}\left( x,y,0\right) \
\chi \left( y\right) \ dy  \label{ns12}
\end{equation}%
that is%
\begin{equation}
\chi \left( y\right) =\frac{c_{1}\beta \left( 0\right) }{2\pi }\int_{-\infty
}^{\infty }K_{h}^{\ast }\left( x,y,0\right) \ \psi \left( x,0\right) \ dx,
\label{ns13}
\end{equation}%
say, by the inverse of the Fourier transform. Thus our equations (\ref{ns11}%
) and (\ref{ns13}) solve the initial value problem for the above linear Schr%
\"{o}dinger equation (\ref{ns10}) as a double integral with the help of the
kernel $K_{h}\left( x,y,t\right) $ that is regular at $t=0,$ when $\mu
\left( 0\right) =c_{1}\neq 0.$

On the other hand,%
\begin{equation}
K_{h}\left( x,y,t\right) =\int_{-\infty }^{\infty }G_{h}\left( x,z,t\right)
\ K_{h}\left( z,y,0\right) \ dz,  \label{ns14}
\end{equation}%
where $G_{h}\left( x,y,t\right) $ is the Green function, which can be obtain
from our solution (\ref{ns2}) in the limit $c_{1}\rightarrow 0$ with a
proper normalization as in the propagator (\ref{in5}). Therefore,
substitution of (\ref{ns13}) into (\ref{ns12}) gives the traditional single
integral form of the solution in terms of the Green function%
\begin{equation}
\psi \left( x,t\right) =\int_{-\infty }^{\infty }G_{h}\left( x,y,t\right) \
\psi \left( y,0\right) \ dy  \label{ns15}
\end{equation}%
by (\ref{ns14}).

The case of the new Hamiltonian, corresponding to (\ref{sol3}), is similar.
The general solution of characteristic equation (\ref{sol14}) is given by%
\begin{eqnarray}
\mu &=&c_{2}\mu _{2}\left( t\right) +c_{3}\mu _{3}\left( t\right)
\label{ns16} \\
&=&\cos t\left( c_{2}\sinh t-c_{3}\cosh t\right) +\sin t\left( c_{2}\cosh
t+c_{3}\sinh t\right)  \notag \\
&=&\frac{1}{\sqrt{2}}\ e^{t}\left( c_{2}\sin \left( t+\frac{\pi }{4}\right)
-c_{3}\cos \left( t+\frac{\pi }{4}\right) \right)  \notag \\
&&-\frac{1}{\sqrt{2}}\ e^{-t}\left( c_{2}\cos \left( t+\frac{\pi }{4}\right)
+c_{3}\sin \left( t+\frac{\pi }{4}\right) \right)  \notag
\end{eqnarray}%
and $\mu ^{\prime }=2\cosh t\left( c_{2}\cos t+c_{3}\sin t\right) $ with $%
\mu \left( 0\right) =-c_{3},$ $\mu ^{\prime }\left( 0\right) =2c_{2}.$ The
first three coefficients of the quadratic form in the solution (\ref{ns2})
are%
\begin{eqnarray}
\alpha \left( t\right) &=&\frac{\cos t\left( c_{2}\cosh t-c_{3}\sinh
t\right) +\sin t\left( c_{2}\sinh t+c_{3}\cosh t\right) }{2\left( \cos
t\left( c_{2}\sinh t-c_{3}\cosh t\right) +\sin t\left( c_{2}\cosh
t+c_{3}\sinh t\right) \right) },  \label{ns17} \\
\beta \left( t\right) &=&\frac{-c_{3}\beta \left( 0\right) }{\cos t\left(
c_{2}\sinh t-c_{3}\cosh t\right) +\sin t\left( c_{2}\cosh t+c_{3}\sinh
t\right) },  \label{ns18} \\
\gamma \left( t\right) &=&\gamma \left( 0\right) +\frac{c_{3}\beta
^{2}\left( 0\right) \left( \cos t\sinh t+\sin t\cosh t\right) }{2\left( \cos
t\left( c_{2}\sinh t-c_{3}\cosh t\right) +\sin t\left( c_{2}\cosh
t+c_{3}\sinh t\right) \right) }  \label{ns19}
\end{eqnarray}%
and one can use formula (\ref{ns9}) for the last coefficient. The
corresponding elementary integral is%
\begin{equation}
\int \frac{dt}{\left( A\cos t+B\sin t\right) ^{2}}=\frac{\sin t}{A\left(
A\cos t+B\sin t\right) }+C.  \label{ns20}
\end{equation}%
The cases (\ref{sol18}) and (\ref{sol19}) can be considered in a similar
fashion. The results are%
\begin{eqnarray}
\mu &=&c_{3}\mu _{3}\left( t\right) +c_{4}\mu _{4}\left( t\right)
\label{ns21} \\
&=&\sin t\left( c_{3}\sinh t+c_{4}\cosh t\right) -\cos t\left( c_{3}\cosh
t+c_{4}\sinh t\right) ,  \notag
\end{eqnarray}%
\begin{eqnarray}
\alpha \left( t\right) &=&\frac{\sin t\left( c_{3}\cosh t+c_{4}\sinh
t\right) +\cos t\left( c_{3}\sinh t+c_{4}\cosh t\right) }{2\left( \sin
t\left( c_{3}\sinh t+c_{4}\cosh t\right) -\cos t\left( c_{3}\cosh
t+c_{4}\sinh t\right) \right) },  \label{ns22} \\
\beta \left( t\right) &=&\frac{-c_{3}\beta \left( 0\right) }{\sin t\left(
c_{3}\sinh t+c_{4}\cosh t\right) -\cos t\left( c_{3}\cosh t+c_{4}\sinh
t\right) },  \label{ns23} \\
\gamma \left( t\right) &=&\gamma \left( 0\right) +\frac{c_{3}\beta
^{2}\left( 0\right) \left( \sin t\cosh t-\cos t\sinh t\right) }{2\left( \sin
t\left( c_{3}\sinh t+c_{4}\cosh t\right) -\cos t\left( c_{3}\cosh
t+c_{4}\sinh t\right) \right) }  \label{ns24}
\end{eqnarray}%
and%
\begin{eqnarray}
\mu &=&c_{1}\mu _{1}\left( t\right) +c_{4}\mu _{4}\left( t\right)
\label{ns25} \\
&=&\cos t\left( c_{1}\cosh t-c_{4}\sinh t\right) +\sin t\left( c_{1}\sinh
t+c_{4}\cosh t\right) ,  \notag
\end{eqnarray}%
\begin{eqnarray}
\alpha \left( t\right) &=&-\frac{\sinh t\left( c_{1}\cos t+c_{4}\sin
t\right) +\cosh t\left( c_{1}\sin t-c_{4}\cos t\right) }{2\left( \sinh
t\left( c_{1}\sin t-c_{4}\cos t\right) +\cosh t\left( c_{1}\cos t+c_{4}\sin
t\right) \right) },  \label{ns26} \\
\beta \left( t\right) &=&\frac{c_{1}\beta \left( 0\right) }{\sinh t\left(
c_{1}\sin t-c_{4}\cos t\right) +\cosh t\left( c_{1}\cos t+c_{4}\sin t\right)
},  \label{ns27} \\
\gamma \left( t\right) &=&\gamma \left( 0\right) +\frac{c_{1}\beta
^{2}\left( 0\right) \left( \cos t\sinh t-\sin t\cosh t\right) }{2\left(
\sinh t\left( c_{1}\sin t-c_{4}\cos t\right) +\cosh t\left( c_{1}\cos
t+c_{4}\sin t\right) \right) },  \label{ns28}
\end{eqnarray}%
respectively. One can use once again formula (\ref{ns9}) for the last
coefficient. We leave further details to the reader.

\section{A Note on The Ill-Posedness of The Schr\"{o}dinger Equations}

The same method shows that the joint solution of the both linear and
nonlinear Schr\"{o}dinger equations (\ref{ns10}) and (\ref{ns1}),
respectively, corresponding to the initial data%
\begin{equation}
\left. \psi \right\vert _{t=0}=\delta _{\varepsilon }\left( x-y\right) =%
\frac{1}{\sqrt{2\pi i\varepsilon }}\ \exp \left( \frac{i\left( x-y\right)
^{2}}{2\varepsilon }\right) ,\quad \varepsilon >0,  \label{ill1}
\end{equation}%
has the form%
\begin{equation}
\psi =G_{\varepsilon }\left( x,y,t\right) =\frac{1}{\sqrt{i\mu _{\varepsilon
}\left( t\right) }}\ e^{i\left( \alpha _{\varepsilon }\left( t\right)
x^{2}+\beta _{\varepsilon }\left( t\right) xy+\gamma _{\varepsilon }\left(
t\right) y^{2}+\kappa _{\varepsilon }\left( t\right) \right) }  \label{ill2}
\end{equation}%
with the characteristic function $\mu _{\varepsilon }\left( t\right) =2\pi
\left( \varepsilon \mu _{1}\left( t\right) +\mu _{2}\left( t\right) \right)
. $ The coefficients of the quadratic form are given by%
\begin{eqnarray}
\alpha _{\varepsilon }\left( t\right) &=&\frac{\cos t\left( \varepsilon
\sinh t+\cosh t\right) -\sin t\left( \varepsilon \cosh t+\sinh t\right) }{%
2\left( \cos t\left( \varepsilon \cosh t+\sinh t\right) +\sin t\left(
\varepsilon \sinh t+\cosh t\right) \right) },  \label{ill3} \\
\beta _{\varepsilon }\left( t\right) &=&-\frac{1}{\cos t\left( \varepsilon
\cosh t+\sinh t\right) +\sin t\left( \varepsilon \sinh t+\cosh t\right) },
\label{ill4} \\
\gamma _{\varepsilon }\left( t\right) &=&\frac{\cos t\cosh t+\sin t\sinh t}{%
2\left( \cos t\left( \varepsilon \cosh t+\sinh t\right) +\sin t\left(
\varepsilon \sinh t+\cosh t\right) \right) }.  \label{ill5}
\end{eqnarray}%
We simply choose $c_{1}=2\pi \varepsilon >0,$ $c_{2}=2\pi $ and $e^{i\varphi
}=1/\sqrt{i}$ and the initial data $\alpha \left( 0\right) =\gamma \left(
0\right) =-\beta \left( 0\right) /2=1/\left( 2\varepsilon \right) $ in a
general solution (\ref{ns6})--(\ref{ns8}). The case $\varepsilon =0,$ $t>0$
corresponds to the original propagator (\ref{in5}), while $\varepsilon >0,$ $%
t=0$ gives the delta sequence (\ref{ill1}).

If $h=h_{\varepsilon }\left( t\right) =\left( \lambda /2\pi \right) \mu
_{\varepsilon }^{\prime }=2\lambda \cos t\left( \varepsilon \sinh t+\cosh
t\right) ,$ then%
\begin{equation}
\kappa _{\varepsilon }\left( t\right) =\left\{
\begin{array}{ll}
-\dfrac{\lambda }{\left( 2\pi \right) ^{s}}\dfrac{\left( \varepsilon \mu
_{1}\left( t\right) +\mu _{2}\left( t\right) \right) ^{1-s}-\varepsilon
^{1-s}}{1-s}, & \text{when }0\leq s<1,\bigskip \\
-\dfrac{\lambda }{2\pi }\ln \left( \mu _{1}\left( t\right) +\dfrac{\mu
_{2}\left( t\right) }{\varepsilon }\right) , & \text{when }s=1%
\end{array}%
\right.  \label{ill6}
\end{equation}%
with $\kappa _{\varepsilon }\left( 0\right) =0$ provided $\varepsilon >0.$

In this example, the initial data $\left. \psi \right\vert
_{t=0}=G_{\varepsilon }\left( x,y,0\right) =\delta _{\varepsilon }\left(
x-y\right) $ converge to the Dirac delta function $\delta \left( x-y\right) $
as $\varepsilon \rightarrow 0^{+}$ in the distributional sense \cite%
{Carleson}, \cite{Sjolin}, \cite{Tao}, \cite{Vega}%
\begin{equation}
\lim_{\varepsilon \rightarrow 0^{+}}\int_{-\infty }^{\infty }G_{\varepsilon
}\left( x,y,0\right) \ \varphi \left( y\right) \ dy=\varphi \left( x\right) .
\label{ill7}
\end{equation}%
On the other hand,%
\begin{eqnarray}
&&\lim_{\varepsilon \rightarrow 0^{+}}\int_{-\infty }^{\infty
}G_{\varepsilon }\left( x,y,t\right) \ \varphi \left( y\right) \ dy
\label{ill7a} \\
&&\qquad =e^{i\lim_{\varepsilon \rightarrow 0^{+}}\kappa _{\varepsilon
}\left( t\right) }\ \int_{-\infty }^{\infty }G_{0}\left( x,y,t\right) \
\varphi \left( y\right) \ dy  \notag
\end{eqnarray}%
with $t>0.$ When $s=1$ the solution $\psi =G_{\varepsilon }\left(
x,y,t\right) ,$ $t>0$ does not have a limit because of divergence of the
logarithmic phase factor $\kappa _{\varepsilon }\left( t\right) $ as $%
\varepsilon \rightarrow 0^{+}.$ See also Refs.~\cite{Ban:Vega} and \cite%
{Ken:Pon:Veg} on the ill-posedness of some canonical dispersive equations.

The second case, corresponding to (\ref{sol3}), is similar. One can choose $%
\mu _{\varepsilon }\left( t\right) =2\pi \left( \mu _{2}\left( t\right)
-\varepsilon \mu _{3}\left( t\right) \right) $ and obtain%
\begin{eqnarray}
\alpha _{\varepsilon }\left( t\right) &=&\frac{\cos t\left( \cosh
t+\varepsilon \sinh t\right) +\sin t\left( \sinh t-\varepsilon \cosh
t\right) }{2\left( \cos t\left( \sinh t+\varepsilon \cosh t\right) +\sin
t\left( \cosh t-\varepsilon \sinh t\right) \right) },  \label{ill8} \\
\beta _{\varepsilon }\left( t\right) &=&-\frac{1}{\cos t\left( \sinh
t+\varepsilon \cosh t\right) +\sin t\left( \cosh t-\varepsilon \sinh
t\right) },  \label{ill9} \\
\gamma _{\varepsilon }\left( t\right) &=&\frac{\cos t\cosh t-\sin t\sinh t}{%
2\left( \cos t\left( \sinh t+\varepsilon \cosh t\right) +\sin t\left( \cosh
t-\varepsilon \sinh t\right) \right) }.  \label{ill10}
\end{eqnarray}%
If $h_{\varepsilon }\left( t\right) =\left( \lambda /2\pi \right) \mu
_{\varepsilon }^{\prime }=2\lambda \cosh t\left( \cos t-\varepsilon \sin
t\right) ,$ then%
\begin{equation}
\kappa _{\varepsilon }\left( t\right) =\left\{
\begin{array}{ll}
-\dfrac{\lambda }{\left( 2\pi \right) ^{s}}\dfrac{\left( \mu _{2}\left(
t\right) -\varepsilon \mu _{3}\left( t\right) \right) ^{1-s}-\varepsilon
^{1-s}}{1-s}, & \text{when }0\leq s<1,\bigskip \\
-\dfrac{\lambda }{2\pi }\ln \left( \dfrac{\mu _{2}\left( t\right) }{%
\varepsilon }-\mu _{3}\left( t\right) \right) , & \text{when }s=1%
\end{array}%
\right.  \label{ill11}
\end{equation}%
and $\kappa _{\varepsilon }\left( 0\right) =0$ when $\varepsilon >0.$
Formulas (\ref{ill2}) and (\ref{ill8})--(\ref{ill11}) describe actual
(nonlinear) evolution for initial data as in (\ref{ill1}). One can observe
once again a discontinuity with respect to these initial data as $%
\varepsilon \rightarrow 0^{+}.$

The cases (\ref{sol18}) and (\ref{sol19}) are as follows. One gets $\mu
_{\varepsilon }\left( t\right) =2\pi \left( \mu _{4}\left( t\right)
-\varepsilon \mu _{3}\left( t\right) \right) ,$%
\begin{eqnarray}
\alpha _{\varepsilon }\left( t\right) &=&\frac{\sin t\left( \sinh
t-\varepsilon \cosh t\right) +\cos t\left( \cosh t-\varepsilon \sinh
t\right) }{2\left( \sin t\left( \cosh t-\varepsilon \sinh t\right) -\cos
t\left( \sinh t-\varepsilon \cosh t\right) \right) },  \label{ill12} \\
\beta _{\varepsilon }\left( t\right) &=&-\frac{1}{\sin t\left( \cosh
t-\varepsilon \sinh t\right) -\cos t\left( \sinh t-\varepsilon \cosh
t\right) },  \label{ill13} \\
\gamma _{\varepsilon }\left( t\right) &=&\frac{\cos t\cosh t-\sin t\sinh t}{%
2\left( \sin t\left( \cosh t-\varepsilon \sinh t\right) -\cos t\left( \sinh
t-\varepsilon \cosh t\right) \right) }  \label{ill14}
\end{eqnarray}%
and $h_{\varepsilon }\left( t\right) =\left( \lambda /2\pi \right) \mu
_{\varepsilon }^{\prime }=2\lambda \sin t\left( \sinh t-\varepsilon \cosh
t\right) ,$%
\begin{equation}
\kappa _{\varepsilon }\left( t\right) =\left\{
\begin{array}{ll}
-\dfrac{\lambda }{\left( 2\pi \right) ^{s}}\dfrac{\left( \mu _{4}\left(
t\right) -\varepsilon \mu _{3}\left( t\right) \right) ^{1-s}-\varepsilon
^{1-s}}{1-s}, & \text{when }0\leq s<1,\bigskip \\
-\dfrac{\lambda }{2\pi }\ln \left( \dfrac{\mu _{4}\left( t\right) }{%
\varepsilon }-\mu _{3}\left( t\right) \right) , & \text{when }s=1%
\end{array}%
\right.  \label{ill15}
\end{equation}%
with $\kappa _{\varepsilon }\left( 0\right) =0,$ $\varepsilon >0$ in the
case (\ref{sol18}). Also $\mu _{\varepsilon }\left( t\right) =2\pi \left(
\mu _{4}\left( t\right) +\varepsilon \mu _{1}\left( t\right) \right) ,$%
\begin{eqnarray}
\alpha _{\varepsilon }\left( t\right) &=&\frac{\sinh t\left( \sin
t+\varepsilon \cos t\right) -\cosh t\left( \cos t-\varepsilon \sin t\right)
}{2\left( \sinh t\left( \cos t-\varepsilon \sin t\right) -\cosh t\left( \sin
t+\varepsilon \cos t\right) \right) },  \label{ill16} \\
\beta _{\varepsilon }\left( t\right) &=&\frac{1}{\sinh t\left( \cos
t-\varepsilon \sin t\right) -\cosh t\left( \sin t+\varepsilon \cos t\right) }%
,  \label{ill17} \\
\gamma _{\varepsilon }\left( t\right) &=&-\frac{\cos t\cosh t+\sin t\sinh t}{%
2\left( \sinh t\left( \cos t-\varepsilon \sin t\right) -\cosh t\left( \sin
t+\varepsilon \cos t\right) \right) }  \label{ill18}
\end{eqnarray}%
and $h_{\varepsilon }\left( t\right) =\left( \lambda /2\pi \right) \mu
_{\varepsilon }^{\prime }=2\lambda \sinh t\left( \sin t+\varepsilon \cos
t\right) ,$%
\begin{equation}
\kappa _{\varepsilon }\left( t\right) =\left\{
\begin{array}{ll}
-\dfrac{\lambda }{\left( 2\pi \right) ^{s}}\dfrac{\left( \mu _{4}\left(
t\right) +\varepsilon \mu _{1}\left( t\right) \right) ^{1-s}-\varepsilon
^{1-s}}{1-s}, & \text{when }0\leq s<1,\bigskip \\
-\dfrac{\lambda }{2\pi }\ln \left( \dfrac{\mu _{4}\left( t\right) }{%
\varepsilon }+\mu _{1}\left( t\right) \right) , & \text{when }s=1%
\end{array}%
\right.  \label{ill19}
\end{equation}%
with $\kappa _{\varepsilon }\left( 0\right) =0,$ $\varepsilon >0$ in the
case (\ref{sol19}). We leave the details to the reader.


\noindent \textbf{Acknowledgments.\/} This paper is written as a part of the
summer 2008 program on analysis of Mathematical and Theoretical Biology
Institute (MTBI) and Mathematical and Computational Modeling Center
(MCMC)=(MC)$^{2}$ at Arizona State University. The MTBI/SUMS Summer
Undergraduate Research Program is supported by The National Science
Foundation (DMS-0502349), The National Security Agency (DOD-H982300710096),
The Sloan Foundation, and Arizona State University.
One of the authors (RCS) is supported by the following National Science Foundation
programs: Louis Stokes Alliances for Minority Participation (LSAMP):
NSF Cooperative Agreement No. HRD-0602425 (WAESO LSAMP Phase IV);
Alliances for Graduate Education and the Professoriate (AGEP):
NSF Cooperative Agreement No. HRD-0450137 (MGE@MSA AGEP Phase II).

\noindent%
The authors are grateful
to Professor Carlos Castillo-Ch\'{a}vez for support and reference \cite%
{Bet:Cin:Kai:Cas}. We thank Professors Alex Mahalov and Svetlana Roudenko
for valuable comments, and Professor Alexander Its for pointing out the
Hamiltonian structure of the characteristic equations; see Appendix~C.

\section{Appendix~A. Fundamental Solutions of The Characteristic Equations}

We denote%
\begin{equation}
u_{1}=\cos t,\qquad u_{2}=\sin t,\qquad v_{1}=\cosh t,\qquad v_{2}=\sinh t
\label{app1}
\end{equation}%
such that $u_{1}^{\prime }=-u_{2},$ $u_{2}^{\prime }=u_{1},$ $v_{1}^{\prime
}=v_{2},$ $v_{2}^{\prime }=v_{1}$ and study differential equations satisfied
by the following set of the Wronskians of trigonometric and hyperbolic
functions%
\begin{equation}
\left\{ W\left( u_{\alpha },v_{\beta }\right) \right\} _{\alpha ,\beta
=1,2}=\left\{ W\left( u_{1},v_{1}\right) ,\ W\left( u_{1},v_{2}\right) ,\
W\left( u_{2},v_{1}\right) ,\ W\left( u_{2},v_{2}\right) \right\} .
\label{app2}
\end{equation}%
Let us take, for example,%
\begin{equation}
y=W\left( u_{1},v_{1}\right) =u_{1}v_{2}+u_{2}v_{1}.  \label{app3}
\end{equation}%
Then%
\begin{equation}
y^{\prime }=2u_{1}v_{1},\qquad y^{\prime \prime }=2u_{1}v_{2}-2u_{2}v_{1}
\label{app4}
\end{equation}%
and%
\begin{equation}
y^{\prime \prime }-\tau y^{\prime }+4\sigma y=\left( 4\sigma +2\right)
u_{1}v_{2}+\left( 4\sigma -2\right) u_{2}v_{1}-2\tau u_{1}v_{1}=0.
\label{app5}
\end{equation}%
The last equation is satisfied when $\sigma =1/2,$ $\tau =2v_{2}/v_{1}$ and $%
\sigma =-1/2,$ $\tau =-2u_{2}/u_{1}.$ All other cases are similar and the
results are presented in Table~1.

Our calculations reveal the following identities%
\begin{eqnarray}
&&W^{\prime \prime }\left( u_{1},v_{1}\right) =-2W\left( u_{2},v_{2}\right)
,\qquad W^{\prime \prime }\left( u_{1},v_{2}\right) =-2W\left(
u_{2},v_{1}\right) ,  \label{app6} \\
&&W^{\prime \prime }\left( u_{2},v_{1}\right) =2W\left( u_{1},v_{2}\right)
,\qquad \ \ W^{\prime \prime }\left( u_{2},v_{2}\right) =2W\left(
u_{1},v_{1}\right) ,  \notag
\end{eqnarray}%
for the Wronskians under consideration. This implies that the set of
Wronskians (\ref{app2}) provides the fundamental solutions of the fourth
order differential equation%
\begin{equation}
W^{\left( 4\right) }+4W=0  \label{app7}
\end{equation}%
with constant coefficients. The corresponding characteristic equation, $%
\lambda ^{4}+4=0,$ has four roots, $\lambda _{1}=1+i,$ $\lambda _{2}=1-i,$ $%
\lambda _{3}=-1+i,$ $\lambda _{4}=-1-i,$ and the fundamental solution set is
given by%
\begin{equation}
\left\{ u_{\alpha }v_{\beta }\right\} _{\alpha ,\beta =1,2}=\left\{
u_{1}v_{1},\ u_{1}v_{2},\ u_{2}v_{1},\ u_{2}v_{2}\right\} .  \label{app8}
\end{equation}%
These solutions of the bi-harmonic equation (\ref{app7}) are even or odd
functions of time. They do not satisfy our second order characteristic
equations. For example, let $w_{1}=u_{1}v_{2}=\cos t\sinh t$ and $%
w_{1}=u_{2}v_{1}=\sin t\cosh t.$ Then, by a direct calculation,%
\begin{eqnarray}
&&L\left( w_{1}\right) =w_{1}^{\prime \prime }+2\tan t\ w_{1}^{\prime
}-2w_{1}=-2\frac{\sinh t}{\cos t},  \label{app9} \\
&&L\left( w_{2}\right) =w_{2}^{\prime \prime }+2\tan t\ w_{2}^{\prime
}-2w_{2}=2\frac{\sinh t}{\cos t}.  \notag
\end{eqnarray}%
Thus, separately, these solutions of (\ref{app7}) satisfy nonhomogeneous
characteristic equations. But together,%
\begin{equation}
L\left( y_{1}\right) =L\left( w_{1}+w_{2}\right) =L\left( w_{1}\right)
+L\left( w_{2}\right) =-2\frac{\sinh t}{\cos t}+2\frac{\sinh t}{\cos t}=0.
\label{app10}
\end{equation}%
A similar property holds for all other solutions of the characteristic
equations from Table~1.\bigskip

\newpage\vfill\eject

\begin{center}
{\large Table~1.} Fundamental solutions of the characteristic
equations.\medskip
\end{center}

\begin{tabular}{|l|l|}
\hline
Characteristic equation $y^{\prime \prime }-\tau y^{\prime }+4\sigma
y=0\qquad $ & Fundamental solution set $\left\{ y_{i},y_{k}\right\}
_{i<k}\qquad $ \\ \hline
$%
\begin{array}{c}
u^{\prime \prime }+u=0 \\
\left( \sigma =1/4,\quad \tau =0\right)%
\end{array}%
$ & $%
\begin{array}{c}
u_{1}=\cos t,\qquad u_{2}=\sin t \\
\left( u_{1}^{\prime }=-u_{2},\quad u_{2}^{\prime }=u_{1}\right)%
\end{array}%
$ \\ \hline
$%
\begin{array}{c}
v^{\prime \prime }-v=0 \\
\left( \sigma =-1/4,\quad \tau =0\right)%
\end{array}%
$ & $%
\begin{array}{c}
v_{1}=\cosh t,\qquad v_{2}=\sinh t \\
\left( v_{1}^{\prime }=v_{2},\quad v_{2}^{\prime }=v_{1}\right)%
\end{array}%
$ \\ \hline
$%
\begin{array}{c}
y^{\prime \prime }+2\tan t\ y^{\prime }-2y=0 \\
\left( \sigma =-1/2,\quad \tau =-2u_{2}/u_{1}\right)%
\end{array}%
$ & $%
\begin{array}{c}
y_{1}=W\left( u_{1},v_{1}\right) =u_{1}v_{2}+u_{2}v_{1} \\
y_{2}=W\left( u_{1},v_{2}\right) =u_{1}v_{1}+u_{2}v_{2}%
\end{array}%
$ \\ \hline
$%
\begin{array}{c}
y^{\prime \prime }-2\cot t\ y^{\prime }-2y=0 \\
\left( \sigma =-1/2,\quad \tau =2u_{2}/u_{1}\right)%
\end{array}%
$ & $%
\begin{array}{c}
y_{3}=W\left( u_{2},v_{2}\right) =u_{2}v_{1}-u_{1}v_{2} \\
y_{4}=W\left( u_{2},v_{1}\right) =u_{2}v_{2}-u_{1}v_{1}%
\end{array}%
$ \\ \hline
$%
\begin{array}{c}
y^{\prime \prime }-2\tanh t\ y^{\prime }+2y=0 \\
\left( \sigma =1/2,\quad \tau =2v_{2}/v_{1}\right)%
\end{array}%
$ & $%
\begin{array}{c}
y_{1}=W\left( u_{1},v_{1}\right) =u_{1}v_{2}+u_{2}v_{1} \\
y_{4}=W\left( u_{2},v_{1}\right) =u_{2}v_{2}-u_{1}v_{1}%
\end{array}%
$ \\ \hline
$%
\begin{array}{c}
y^{\prime \prime }-2\coth t\ y^{\prime }+2y=0 \\
\left( \sigma =1/2,\quad \tau =2v_{1}/v_{2}\right)%
\end{array}%
$ & $%
\begin{array}{c}
y_{2}=W\left( u_{1},v_{2}\right) =u_{1}v_{1}+u_{2}v_{2} \\
y_{3}=W\left( u_{2},v_{2}\right) =u_{2}v_{1}-u_{1}v_{2}%
\end{array}%
$ \\ \hline
\end{tabular}

In order to obtain the fundamental solutions in an algebraic manner, we
denote%
\begin{eqnarray}
&&L_{1}=\dfrac{d^{2}}{dt^{2}}+2\frac{u_{2}}{u_{1}}\dfrac{d}{dt}-2,\quad
L_{2}=\frac{d^{2}}{dt^{2}}-2\frac{u_{1}}{u_{2}}\frac{d}{dt}-2,  \label{app11}
\\
&&L_{3}=\frac{d^{2}}{dt^{2}}-2\frac{v_{2}}{v_{1}}\frac{d}{dt}+2,\quad L_{4}=%
\frac{d^{2}}{dt^{2}}-2\frac{v_{1}}{v_{2}}\frac{d}{dt}+2  \notag
\end{eqnarray}%
and compute the actions of these second order linear differential operators $%
L_{k}$ on the four basis vectors $\left\{ u_{\alpha }v_{\beta }\right\}
_{\alpha ,\beta =1,2},$ namely, $L_{k}\left( u_{\alpha }v_{\beta }\right) .$
The results are presented in Table~2.

\begin{center}
{\large Table~2.} Construction of the fundamental solutions.\medskip

\begin{tabular}{|l|l|l|l|l|}
\hline
Linear operators & $u_{1}v_{1}$ & $u_{1}v_{2}$ & $u_{2}v_{1}$ & $u_{2}v_{2}$
\\ \hline
$%
\begin{array}{c}
\dfrac{d}{dt}%
\end{array}%
\begin{array}{c}
\  \\
\
\end{array}%
$ & $u_{1}v_{2}-u_{2}v_{1}$ & $u_{1}v_{1}-u_{2}v_{2}$ & $%
u_{1}v_{1}+u_{2}v_{2}$ & $u_{1}v_{2}+u_{2}v_{1}$ \\ \hline
$%
\begin{array}{c}
\dfrac{d^{2}}{dt^{2}}%
\end{array}%
\begin{array}{c}
\  \\
\
\end{array}%
$ & $-2u_{2}v_{2}$ & $-2u_{2}v_{1}$ & $2u_{1}v_{2}$ & $2u_{1}v_{1}$ \\ \hline
$%
\begin{array}{c}
L_{1}%
\end{array}%
\begin{array}{c}
\  \\
\
\end{array}%
$ & $-2\dfrac{v_{1}}{u_{1}}$ & $-2\dfrac{v_{2}}{u_{1}}$ & $2\dfrac{v_{2}}{%
u_{1}}$ & $2\dfrac{v_{1}}{u_{1}}$ \\ \hline
$%
\begin{array}{c}
L_{2}%
\end{array}%
\begin{array}{c}
\  \\
\
\end{array}%
$ & $-2\dfrac{v_{2}}{u_{2}}$ & $-2\dfrac{v_{1}}{u_{2}}$ & $-2\dfrac{v_{1}}{%
u_{2}}$ & $-2\dfrac{v_{2}}{u_{2}}$ \\ \hline
$%
\begin{array}{c}
L_{3}%
\end{array}%
\begin{array}{c}
\  \\
\
\end{array}%
$ & $2\dfrac{u_{1}}{v_{1}}$ & $-2\dfrac{u_{2}}{v_{1}}$ & $2\dfrac{u_{2}}{%
v_{1}}$ & $2\dfrac{u_{1}}{v_{1}}$ \\ \hline
$%
\begin{array}{c}
L_{4}%
\end{array}%
\begin{array}{c}
\  \\
\
\end{array}%
$ & $2\dfrac{u_{2}}{v_{2}}$ & $-2\dfrac{u_{1}}{v_{2}}$ & $-2\dfrac{u_{1}}{%
v_{2}}$ & $-2\dfrac{u_{2}}{v_{2}}$ \\ \hline
\end{tabular}%
\medskip
\end{center}

Therefore%
\begin{eqnarray}
&&L_{1}\left( u_{1}v_{1}+u_{2}v_{2}\right) =L_{1}\left(
u_{1}v_{2}+u_{2}v_{1}\right)  \label{app12} \\
&&\quad =L_{2}\left( u_{2}v_{2}-u_{1}v_{1}\right) =L_{2}\left(
u_{2}v_{1}-u_{1}v_{2}\right)  \notag \\
&&\quad =L_{3}\left( u_{1}v_{2}+u_{2}v_{1}\right) =L_{3}\left(
u_{2}v_{2}-u_{1}v_{1}\right)  \notag \\
&&\quad =L_{4}\left( u_{1}v_{1}+u_{2}v_{2}\right) =L_{4}\left(
u_{2}v_{1}-u_{1}v_{2}\right) =0  \notag
\end{eqnarray}%
as has been stated in Table~1.

All our characteristic equations in this paper obey certain periodicity
properties. For instance, equations%
\begin{equation}
y^{\prime \prime }+2\tan t\ y^{\prime }-2y=0  \label{app13}
\end{equation}%
and%
\begin{equation}
y^{\prime \prime }-2\cot t\ y^{\prime }-2y=0  \label{app14}
\end{equation}%
are invariant under the shifts $t\rightarrow t\pm \pi $ and interchange one
into another when $t\rightarrow t\pm \pi /2.$ Since only two solutions of a
linear second order differential equation may be linearly independent, the
corresponding fundamental solutions satisfy the following relations%
\begin{equation}
\left(
\begin{array}{c}
y_{1}\left( t\pm \pi \right) \\
y_{2}\left( t\pm \pi \right)%
\end{array}%
\right) =-\left(
\begin{array}{cc}
\cosh \pi & \pm \sinh \pi \\
\pm \sinh \pi & \cosh \pi%
\end{array}%
\right) \left(
\begin{array}{c}
y_{1}\left( t\right) \\
y_{2}\left( t\right)%
\end{array}%
\right)  \label{app15}
\end{equation}%
and%
\begin{equation}
\left(
\begin{array}{c}
y_{1}\left( t\pm \pi /2\right) \\
y_{2}\left( t\pm \pi /2\right)%
\end{array}%
\right) =-\left(
\begin{array}{cc}
\sinh \left( \pi /2\right) & \pm \cosh \left( \pi /2\right) \\
\pm \cosh \left( \pi /2\right) & \sinh \left( \pi /2\right)%
\end{array}%
\right) \left(
\begin{array}{c}
y_{3}\left( t\right) \\
y_{4}\left( t\right)%
\end{array}%
\right) ,  \label{app16}
\end{equation}%
respectively. Two other characteristic equations have pure imaginary
periods. We leave the details to the reader.

\section{Appendix~B. On A Transformation of The Quantum Hamiltonians}

Our definition of the creation and annihilation operators given by (\ref{in3}%
) implies the following operator identities%
\begin{equation}
x^{2}=\frac{1}{2}\left( aa^{\dagger }+a^{\dagger }a\right) -\frac{1}{2}%
\left( a^{2}+\left( a^{\dagger }\right) ^{2}\right) ,  \label{appb1a}
\end{equation}%
\begin{equation}
\frac{\partial ^{2}}{\partial x^{2}}=-\frac{1}{2}\left( aa^{\dagger
}+a^{\dagger }a\right) -\frac{1}{2}\left( a^{2}+\left( a^{\dagger }\right)
^{2}\right) ,  \label{appb1b}
\end{equation}%
\begin{equation}
2x\frac{\partial }{\partial x}+1=-a^{2}+\left( a^{\dagger }\right) ^{2}
\label{appb1c}
\end{equation}%
(and vise versa), which allows us to transform the time-dependent Schr\"{o}%
dinger equation (\ref{sol1}) into a Hamiltonian form (\ref{in1}), where the
Hamiltonian is written in terms of the creation and annihilation operators
as follows%
\begin{eqnarray}
H &=&\frac{1}{2}\left( a\left( t\right) +b\left( t\right) \right) \left(
aa^{\dagger }+a^{\dagger }a\right)  \label{appb2} \\
&&+\frac{1}{2}\left( a\left( t\right) -b\left( t\right) +2id(t)\right) a^{2}+%
\frac{1}{2}\left( a\left( t\right) -b\left( t\right) -2id(t)\right) \left(
a^{\dagger }\right) ^{2},  \notag
\end{eqnarray}%
when $c=2d.$ This helps to transform the Hamiltonians of modified
oscillators under consideration into different equivalent forms, which are
used in the paper.

The trigonometric cases (\ref{sol2}) and (\ref{sol18}) results in the
Hamiltonians (\ref{in2}) and (\ref{nd14}) with $n=1,$ respectively. In the
first hyperbolic case (\ref{sol3}) one gets%
\begin{equation}
H=\frac{1}{2}\cosh \left( 2t\right) \left( aa^{\dagger }+a^{\dagger
}a\right) +\frac{1}{2}\left( 1-i\sinh \left( 2t\right) \right) a^{2}+\frac{1%
}{2}\left( 1+i\sinh \left( 2t\right) \right) \left( a^{\dagger }\right) ^{2},
\label{appb3}
\end{equation}%
where%
\begin{equation}
1\pm i\sinh \left( 2t\right) =\cosh \left( 2t\right) e^{\pm i\arctan \left(
2\tau \right) },\qquad \tau =\frac{1}{2}\sinh \left( 2t\right) ,
\label{appb4}
\end{equation}%
which implies the Schr\"{o}dinger equation (\ref{in6})--(\ref{in7}). The
second hyperbolic case (\ref{sol19}) is similar. We leave the details to the
reader.

\section{Appendix C. On A Hamiltonian Structure of The Characteristic
Equations}

The Hamilton equations of classical mechanics \cite{Lan:Lif},%
\begin{equation}
\overset{.}{q}=\frac{\partial H}{\partial p},\qquad \overset{.}{p}=-\frac{%
\partial H}{\partial q},  \label{appc1}
\end{equation}%
with a general quadratic Hamiltonian%
\begin{equation}
H=a\left( t\right) p^{2}+b\left( t\right) q^{2}+2d\left( t\right) pq
\label{appc2}
\end{equation}%
are%
\begin{equation}
\overset{.}{q}=2ap+2dq,\qquad \overset{.}{p}=-2bq-2dp.  \label{appc3}
\end{equation}%
We denote, as is customary, differentiation with respect to time by placing
a dot above the canonical variables $p$ and $q.$ Elimination of the
generalized momentum $p$ from this system results in the second order
equation with respect to the generalized coordinate%
\begin{equation}
\overset{..}{q}-\frac{\overset{.}{a}}{a}\ \overset{.}{q}+4\left( ab-d^{2}+%
\frac{d}{2}\left( \frac{\overset{.}{a}}{a}-\frac{\overset{.}{d}}{d}\right)
\right) q=0.  \label{appc4}
\end{equation}%
It coincides with the characteristic equation (\ref{sol8})--(\ref{sol9})
with $c=2d.$ Our choice of the coefficients (\ref{sol2})--(\ref{sol3}) and (%
\ref{sol18})--(\ref{sol19}) in the classical Hamiltonian (\ref{appc2})
corresponds to the following models of modified classical oscillators%
\begin{eqnarray}
\overset{..}{q}+2\tan t\ \overset{.}{q}-2q &=&0,  \label{ahc1} \\
\overset{..}{q}-2\tanh t\ \overset{.}{q}+2q &=&0,  \label{ahc2} \\
\overset{..}{q}-2\cot t\ \overset{.}{q}-2q &=&0,  \label{ahc3} \\
\overset{..}{q}-2\coth t\ \overset{.}{q}+2q &=&0,  \label{ahc4}
\end{eqnarray}%
respectively; see Appendix~A for their fundamental solutions.

The standard quantization of the classical integrable systems under
consideration, namely,%
\begin{equation}
q\rightarrow x,\qquad p\rightarrow i^{-1}\frac{\partial }{\partial x},\qquad %
\left[ x,p\right] =xp-px=i  \label{appc5}
\end{equation}%
and%
\begin{equation}
H\rightarrow ap^{2}+bx^{2}+d\left( px+xp\right) ,\qquad i\frac{\partial \psi
}{\partial t}=H\psi ,  \label{appc6}
\end{equation}%
leads to the quantum exactly solvable models of modified oscillators
discussed in this paper.

Another example is a damped oscillator with $a=\left( \omega _{0}/2\right)
e^{-2\lambda t},$ $b=\left( \omega _{0}/2\right) e^{2\lambda t}$ and $c=d=0.$
The classical equation%
\begin{equation}
\overset{..}{q}+2\lambda \ \overset{.}{q}+\omega _{0}^{2}\ q=0  \label{appc7}
\end{equation}%
describes damped oscillations \cite{Lan:Lif}. The corresponding quantum
propagator has the form (\ref{sol4}) with%
\begin{equation}
\mu =\frac{\omega _{0}}{\omega }e^{-\lambda t}\sin \omega t,\qquad \omega
^{2}=\omega _{0}^{2}-\lambda ^{2}>0  \label{appc8}
\end{equation}%
and%
\begin{eqnarray}
\alpha \left( t\right) &=&\frac{\omega \cos \omega t-\lambda \sin \omega t}{%
2\omega _{0}\sin \omega t}e^{2\lambda t},  \label{appc9} \\
\beta \left( t\right) &=&-\frac{\omega }{\omega _{0}\sin \omega t}e^{\lambda
t},  \label{appc10} \\
\gamma \left( t\right) &=&\frac{\omega ^{2}-\omega _{0}^{2}\sin ^{2}\omega t%
}{2\omega _{0}\sin \omega t\left( \omega \cos \omega t-\lambda \sin \omega
t\right) }.  \label{appc11}
\end{eqnarray}%
The Schr\"{o}dinger equation%
\begin{equation}
i\frac{\partial \psi }{\partial t}=\frac{\omega _{0}}{2}\left( -e^{-2\lambda
t}\frac{\partial ^{2}\psi }{\partial x^{2}}+e^{2\lambda t}x^{2}\psi \right)
\label{appc12}
\end{equation}%
describes the linear oscillator with a variable unit of length $x\rightarrow
xe^{\lambda t}.$ See \cite{Cor-Sot:Sua:Sus} for more details.

\end{document}